\documentclass[twocolumn]{aastex631}
\usepackage[utf8]{inputenc}
\usepackage{threeparttable}

\defcitealias{weaver2022a}{W22}

\begin{document}

\title{A Value-added COSMOS2020 Catalog of Physical Properties:  Constraining Temperature-dependent Initial Mass Function}

\correspondingauthor{Vadim Rusakov}
\email{rusakov124@gmail.com}

\author[0000-0001-7633-3985]{Vadim Rusakov}
\affiliation{Cosmic Dawn Center (DAWN)}
\affiliation{Niels Bohr Institute, University of Copenhagen, Lyngbyvej 2, DK-2100 Copenhagen \O}

\author[0000-0003-3780-6801]{Charles L. Steinhardt}
\affiliation{Cosmic Dawn Center (DAWN)}
\affiliation{Niels Bohr Institute, University of Copenhagen, Lyngbyvej 2, DK-2100 Copenhagen \O}

\author[0000-0002-5460-6126]{Albert Sneppen}
\affiliation{Cosmic Dawn Center (DAWN)}
\affiliation{Niels Bohr Institute, University of Copenhagen, Lyngbyvej 2, DK-2100 Copenhagen \O}

\begin{abstract}

% Abstract 1: Presenting the catalog
This work presents and releases a catalog of new photometrically-derived physical properties for the $\sim 10^5$ most well-measured galaxies in the COSMOS field on the sky.  Using a recently developed technique, spectral energy distributions are modeled assuming a stellar initial mass function (IMF) that depends on the temperature of gas in star-forming regions.  The method is applied to the largest current sample of high-quality panchromatic photometry, the COSMOS2020 catalog, that allows for testing this assumption.  It is found that the galaxies exhibit a continuum of IMF, and gas temperatures, most of which are bottom-lighter than measured in the Milky Way.  As a consequence, the stellar masses and star formation rates of most galaxies here are found to be lower than those measured by traditional techniques in the COSMOS2020 catalog by factors of $\sim1.6-3.5$ and $2.5-70.0$, respectively, with the change being the strongest for the most active galaxies.  The resulting physical properties provide new insights into variation of the IMF-derived gas temperature along the star-forming main sequence and at quiescence, produce a sharp and coherent picture of downsizing, as seen from the stellar mass functions, and hint at a possible high-temperature and high-density stage of early galactic evolution.

\end{abstract}


\section{Introduction} \label{sec:intro}

Most galaxies outside of the Local Group are studied by fitting photometric templates with known physical properties.  These templates rely on strong assumptions about the formation and evolution of stellar populations \citep{conroy2013}.  The problem is complicated, as most of the starlight is integrated and comes from the high-mass stars that compose a small fraction of its total stellar mass.  Typically in models, the stellar population is produced by some form of a Galactic stellar initial mass function (IMF; such as \citealp{salpeter1955,miller1979,kroupa2001,chabrier2003}) combined with a specific star formation history and dust extinction.  Although the IMF is constrained fairly well in the locality of the Milky Way (MW), where it is possible to resolve individual stars, the same task is impossible for distant enough galaxies.  Therefore, as this assumption fixes a certain stellar budget for galaxies, it is crucial that it holds so that the physical properties are estimated accurately.  Otherwise, it can break the results and the interpretation of different stages of galaxy evolution.

Several observations have hinted that models of integrated starlight calibrated against stellar populations in the local universe may sometimes fail at increasingly high redshift.  
Large galaxy surveys \citep{hildebrandt2009,caputi2015,finkelstein2015,steinhardt2016} have reported that dark matter halos inferred from stellar masses at high redshift become too massive to fit cosmological predictions.  Furthermore, recent observations of galaxies at $z>8$ (\citealp{atek2023,labbe2023,naidu2022}, among others) have revealed that by drawing from the same stellar budget they require more baryons than there are produced at that time \citep{boylankolchin2023}.  At the same time, the observed luminosity functions have been under-predicted in some of the semi-analytical simulations at $z>13$ by a factor of 30\footnote{ It is worth mentioning that other studies have not reported the tension with interpreted or simulated galaxy properties (\citealp{adams2023,mccaffrey2023}).  Others have indicated that it can be resolved by changing assumptions in galaxy models \citep{yung2023}.
 Therefore, the apparent nature of the tension reinforces the fact that it is likely a test of the galaxy models and not the cosmology.}.

On the other hand, low-redshift observations sensitive to low-mass stars hint at underestimated stellar mass in early-type galaxies.  For example, measurements of stellar kinematics, absorption lines or various line indices find excess in mass-to-light ratio in local elliptical galaxies relative to the Milky Way IMF \citep{cappellari2012,conroy_vandokkum2012,vandokkum_conroy2012,martinnavarro2015,vandokkum_conroy2017}.  Other observations of resolved stellar populations in the MW or local dwarfs demonstrate similar IMF results in old stellar structures \citep{geha2013,hallakoun2021}.

Together, these pieces of evidence independently suggest that the mass-to-light ratio in galaxy models appears to be off in the edge cases with drastically different star-forming conditions.  It is too high for active galaxies at high redshift and too conservative for passive ones at low redshift.  At the same time the galaxy populations with intermediate properties likely represent the bulk of the distribution.  

In the high redshift cases where the tension with the cosmology arises, \cite{steinhardt2022c_templates,harikane2023} suggested that a bottom-lighter (or top-heavier) IMF can account for the over-estimation of stellar mass inferred from observations.  \cite{jermyn2018,steinhardt2022c_templates} proposed that the bottom-light IMF at high redshift is likely driven by the cosmic microwave background (CMB) temperature.  Similarly, \cite{yung2023} proposed that the simulations can match the higher galaxy number densities with a top-heavier IMF.  It was also possible to resolve the tension by more properly accounting for cosmic variance, uncertainty in the stellar mass and accounting for possible additional gas for star formation \citep{chen2023}.

Consequently, one could also argue for a more bottom-heavy IMF to solve the tension on the other redshift end (in the local universe) to account for the reported under-estimation of stellar mass in the early-type galaxies.  This would be in agreement with the IMF measurements of \cite{cappellari2012,conroy_vandokkum2012,vandokkum_conroy2012,martinnavarro2015,vandokkum_conroy2017}.  Therefore, if the shape of the IMF varies with average conditions of star formation in different galaxies, there is likely to be a continuum of stellar mass functions.

In theory, changes in temperature, density or metallicity of the star-forming material must lead to a corresponding change in the relevant mass scales of the stellar population \citep{jeans1902,low1976,larson1998,larson2005} and modify the shape of the IMF \citep{bate2005,jappsen2005,krumholz2011,hopkins2012,jermyn2018}.  Although, the star-forming gas conditions have been difficult to determine from line ratios at high redshift, the temperature of cold dust serves as a strong indicator of conditions of the interstellar medium (ISM).  A possible link between the thermal state of the gas and dust in the main-sequence galaxies has been shown by comparing mass and luminosity-weighted dust temperatures with excitation temperature from [C {\sc i}] line ratio \citep{valentino2020}.

Empirically, measurements of dust temperature hint that the state of the ISM (and possibly the star-forming gas) changes with redshift and position with respect to the star-forming main sequence (SFMS; \citealp{magnelli2014,lamperti2019}).  The latter studies show that the temperature of the cold component of dust measured from the far-IR (FIR) continuum increases from $\sim 20$ to 40 K for galaxies depending on the offset from the SFMS.  Besides, the average dust temperature increases with redshift \citep{schreiber2018,cortzen2020}, which is similar to the reported increase in the intensity of the radiation field at higher redshift \citep{bethermin2015,magdis2017}.  If the molecular clouds exhibit similar behavior, this evidence builds on to suggest that the IMF should change accordingly \citep{kroupa2001,bastian2010}. 

Moreover, \cite{gunawardhana2011,nanayakkara2017} show that the excessive equivalent widths of the $H \alpha$ line are likely due to a top-heavier IMF at $z \sim 0.35$ and $z \sim 2$.  As this excess correlates with the star formation rate, they suggest that the effect increases with redshift for star-forming galaxies.  In addition, a more direct probe of the IMF in \cite{zhang2018} -- $^{13}{\rm C}/^{18}{\rm O}$ transitions -- reveals a top-heavier IMF in several starbursts at $2<z<3$.

The recently introduced technique in \cite{sneppen2022} was used to constrain the IMF in some of the most well-measured galaxies.  However, it has been challenging to test the idea due to rare photometric band coverage and relatively low quality of available measurements.  Therefore, most samples are prone to the strong covariances between various model parameters.
As such, the lack of emission from the bulk of stellar mass leads to degeneracies between the IMF, the law of dust attenuation or the star formation history.  However, \cite{sneppen2022} showed that improved measurements in the recent data sets (such as COSMOS2015; \citealp{laigle2016}) can provide more statistical power to relax the assumption of the local IMF and break the existing degeneracies to some extent.

This work employs the fitting technique to produce a catalog of new physical properties for galaxies in the COSMOS field, available for use by the community.  It is expected that new higher quality observations released in the COSMOS2020 catalog \citep{weaver2022a} can reveal fainter galaxies across a range of redshifts, as well as refine the observations of the previous catalogs.  With these data, it is possible to improve on the estimates of stellar mass ($M_\star$) and star formation rate (SFR), by starting with the, possibly, more physically plausible set of assumptions for the stellar IMF.  In addition, this update provides the IMF property theoretically related to the temperature of the star-forming gas \citep{jermyn2018} and argues for its importance for galaxy evolution.  The work here shows that the measured change in stellar mass and star formation rate and the associated increase in the inferred gas temperature at lower redshifts ($0 < z < 2$) reinforces the importance of the proper treatment of stellar populations particularly at high redshift, where the cosmological predictions can be tested in the most direct way.

The data used in this work are described in \S~\ref{sec:data}.  Next, \S~\ref{sec:sed_fitting} outlines the procedure of SED fitting, which includes the description of the temperature-dependent IMF and the associated templates.  Then, \S~\ref{sec:results} details the quality choices made to produce the final data set and demonstrates the best-fit physical properties.  The following subsections demonstrate some well-known relationships with the new properties and describe new insights driven by the best-fit IMF.  Finally, the discussion in \S~\ref{sec:discussion} reiterates some of the most critical approximations made when interpreting observations in application to galaxy evolution.  This work assumes a Flat Lambda Cold Dark Matter cosmology ($H_0 = 70$ km s$^{-1}$ Mpc$^{-1}$, $\Omega_{m} = 0.3$, $\Omega_{\Lambda}=0.7$).

\section{Data catalog} \label{sec:data}

Previously, work to constrain the IMF using this technique was done with the COSMOS2015 catalog \citep{laigle2016} in \cite{sneppen2022}, which when published was the largest existing multi-band photometric data set.  COSMOS2020 \citep{weaver2022a} is an updated catalog with more detections and deeper and more precise photometry than the previous versions in the COSMOS field \citep{capak2007,scoville2007,ilbert2009,ilbert2013,laigle2016}.  The most significant advantages arise from deeper infrared bands and from including bluer bands in its detection image.  The latter is expected to result in a higher number of small blue galaxies, which is particularly useful for this work, where compact blue galaxies are of special interest, as discussed in \S~\ref{sec:hrdiagram}.

There are two primary versions of the COSMOS2020 catalog, \texttt{CLASSIC} and \texttt{FARMER}.  This work makes use of the \texttt{CLASSIC} version, which is preferred for the brightest galaxies.  In contrast to the modeled fluxes in the \texttt{FARMER} version, it derives aperture-based photometry from PSF-homogenized images similarly to the previous catalogs.  Despite the differences in approach, the precision of measurements and the fraction of outliers in terms of photometric redshifts are similar between them on average.  For the brightest galaxies \texttt{CLASSIC} performs marginally better both in precision and number of redshift outliers using the standard photometric fitting procedure.  The \texttt{FARMER} catalog is analysed in a similar experiment with a different photometric fitting code, {\texttt LePhare} \citep{arnouts1999,ilbert2006} in Rusakov et al. (2023; in prep.).

The photometric bands were selected as per the suggestions in \cite{weaver2022a}.  Several filters were discarded as too shallow and thus not providing additional constraints: Spitzer/IRAC Channels 3 and 4; Subaru Suprime-Cam (SC) broad bands, $NB711$, $NB816$, $NB118$; and GALEX FUV and NUV.  The final list of photometric bands included 25 filters: Canada France Hawaii Telescope (CFHT) $u$, $u^*$; five Subaru/Hyper Suprime-Cam (HSC) bands $grizy$; four UltraVISTA DR4 bands $YJHK_S$; twelve Subaru/Suprime-Cam (SC) bands $IB427$, $IB464$, $IA484$, $IB505$, $IA527$, $IB574$, $IA624$, $IA679$, $IB709$, $IA738$, $IA767$, $IB827$; and Spitzer/IRAC channels 1, 2.  The work makes use of the deeper fluxes derived with the 2-arcsecond apertures.  The photometry was corrected for aperture sizes and the Milky Way extinction based on the offsets provided in the catalog.

To obtain the most accurate and well-measured objects in the COSMOS2020 catalog, several selections were made.  Of the 1,720,700 sources, approximately 437,000 were kept by removing the objects that were flagged by \texttt{SExtractor}, including sources biased by the bright neighbors, at the boundaries of the images, saturated detections or instances that failed at execution.  These sources were selected in ``ultradeep'' stripes of the UltraVISTA photometry, which reduced the survey area from $\sim3.4$ to 0.9 deg$^2$ and decreases the volume completeness of the final catalog.
Further selections of objects were made after solving for SED templates with the best-fit IMF, as some of the poor measurements were found to be most prone to degeneracies in the solutions and ended up at the edges of the IMF temperature range.  This quality cut is described in \S~\ref{sec:quality_cuts}.

\section{SED Fitting with \texttt{EAZY}} \label{sec:sed_fitting}

The algorithm for fitting photometric SEDs with \texttt{EAZY} \citep{brammer2008,eazypy2021} is based on a linear combination of SED templates from a model set.  This procedure significantly reduces the computation time, as the best-fit linear combination can be determined via matrix inversion rather than via extended grid sampling.  

However, if the set of basis templates is not carefully constructed, the best-fit linear combination may produce an nonphysical shortcut solution.  This is possible as almost all of the resulting solutions are represented by a linear mixture of young and old stellar populations with varying amounts of dust.  Even if the problem is well-determined, statistically, the ultraviolet and infrared components of observed SEDs can be reconstructed with different combinations of dust, star formation history and the shape of the IMF.  Thus, for example, a quiescent galaxy at may be fitted with an SED of a dusty star-forming galaxy at lower redshift, unless the strong emission lines can be sampled by a fortunate coincidence of some narrow bands.

Therefore, it is crucial to construct a set of model templates to match the prior expectations of physical properties of galaxies in the targeted redshift regime.  The templates are typically made to span a limited and physically-motivated space of such parameters, as stellar metallicity, age and dust extinction.  On the other hand, the spanned space is also made substantially large to sample the properties as much as possible and avoid forcing most of the inferred galaxy properties to look the same.  The balance between asserting a narrow physical domain of templates and still obtaining a representative fit is achieved by ensuring that the template galaxies efficiently span the color space \citep{brammer2008}.  This work builds on the templates provided for COSMOS2020 at the time of publication\footnote{Standard COSMOS2020 templates made with a Chabrier IMF are available at: \url{https://github.com/gbrammer/eazy-photoz/}.}.

\subsection{Temperature-dependent IMF} \label{sec:imf}

It is likely that different phases of gas in the giant molecular clouds result in IMF variability across individual galaxies at different physical locations and at different times of their evolution.  Various dependencies of the IMF on the gas phase have been proposed \citep{hopkins2012,krumholz2011,jappsen2005,bate2005}, where some of the simplest approaches use the scaling of the IMF mass with the temperature only \citep{jermyn2018}.  The mass scale of cloud fragmentation in the Kroupa IMF \citep{kroupa2001} is made proportional to the square of the temperature:

\begin{equation} \label{eq:imf}
    \frac{dN}{dm}(T) \propto 
    \left\{
    \begin{array}{ll}
        m^{-0.3}, & m < 0.08 M_{\odot} f(T) \\ 
        m^{-1.3}, & 0.08 M_{\odot} f(T) < m < 0.50 M_{\odot} f(T) \\ 
        m^{-2.3}, & 0.50 M_{\odot} f(T) < m
    \end{array}
    \right.,
\end{equation}

where $f(T) = (T/T_0)^2$ is the scaling factor and $T_0 = 20$ K is the reference temperature set to approximately the temperature of the molecular clouds in the Milky Way \citep{schnee2008}.  The scaling of the power-law break-points is tied to the minimum fragmentation mass of a gas cloud, i.e. the low-mass end of the IMF \citep{jermyn2018}.  Therefore, throughout this paper we refer to the relative IMF changes as bottom-light or bottom-heavy, although in practice the bottom-light(heavy) and top-heavy(light) shapes are degenerate here, as the power-law slopes are fixed.

Although it is unclear precisely how the complex interplay of various astrophysical mechanisms in molecular clouds sets their exact temperature, perhaps, a minimum at $z > 7$ is set by the CMB, where it exceeds the local temperature of 20 K.  At lower redshifts, where the temperature exceeds the microwave background, as for the $\sim 20$ K in star-formirng regions of the Milky Way, it could be regulated by stellar radiation in regular main-sequence galaxies or by cosmic rays in stellar explosions \citep{papadopoulos2010}.

Although theoretically, the IMF is parameterised using the gas temperature, it is unclear whether the best-fit IMF constrains this quantity or is degenerate with other model SED inputs.  For example, it was argued by \cite{steinhardt2022_III} that the SED templates with a parametrised IMF are likely sensitive to either the sharp decline in the star formation history or just the IMF as the first order effect.  There is less confidence that either of these can directly translate to the temperature of the molecular gas.  However, the star-forming main sequence is found to correlate with the IMF similarly to the temperature of dust with redshift in \cite{magnelli2014}, which lends more credence to the IMF-temperature interpretation.  Moreover, the connection of the IMF temperature to galaxy morphologies in \cite{steinhardt2023} further reinforces the physical origin of this property.

Photometric templates used in this work were constructed to have as similar properties as possible to the standard COSMOS2020 templates\footnote{The standard templates preserved the records of their parameters, except for the star formation history.  Therefore it was adjusted to produce as similar templates here as possible.}, with the sole exception of changing the IMF.  The standard 17 templates span stellar populations with ages from 0.1 to 7 Gyr, dust extinction ranging $\sim0 - 2$ magnitudes, constant star formation histories with specific star formation rate $sSFR \sim 2 \times 10^{-8}$ yr$^{-1}$ and the attenuation curve from \cite{kriek2013} (Table~\ref{tab:templates}).  Building upon this 17-template basis, a series of templates were made with the varying IMF shapes that corresponded to molecular gas temperatures ranging from 10 to 60 K in steps of 1 K.  They were produced using \texttt{Flexible Stellar Population Synthesis} (\texttt{FSPS}) code \citep{conroy2009,conroy2010} and made available publicly\footnote{The templates with a range of varying IMFs are made publicly available at \href{https://github.com/vvrus/sed-templates}{https://github.com/vvrus/sed-templates}.}.

\begin{table}
\centering
    \caption{Properties of the stellar populations in the 17-template basis of standard templates used with COSMOS2020. The columns show: ID of a template; age of a stellar population; fraction of a solar metallicity $log(Z/Z_{\odot})$; and dust extinction in the V band, $A_V$.  The first 7 templates are produced using Padova isochrones with $Z_{\odot}=0.019$ and for the last 10 MIST isochrones that assume $Z_{\odot}=0.0142$ were used. }
    \label{tab:templates}
    \begin{tabular}{cccc}
        \hline\hline
        ID & Age (Gyr) & $log(Z/Z_{\odot})$ & $A_{V}$  \\ \hline
        1  & 0.1  & 0.0 & 0.005 \\
        2  & 0.1  & 0.0 & 0.5   \\
        3  & 0.1  & 0.0 & 0.005 \\
        4  & 0.1  & 0.0 & 0.5   \\
        5  & 0.1  & 0.0 & 1.0   \\
        6  & 0.1  & 0.0 & 2.0   \\
        7  & 0.1  & 0.0 & 3.0   \\
        8  & 0.31 & 0.0 & 0.005 \\
        9  & 0.31 & 0.0 & 1.0   \\
        10 & 0.31 & 0.0 & 2.0   \\
        11 & 0.62 & 0.0 & 0.005 \\
        12 & 0.62 & 0.0 & 1.0   \\
        13 & 0.62 & 0.0 & 2.5   \\
        14 & 1.76 & 0.0 & 0.005 \\
        15 & 1.76 & 0.0 & 1.0   \\
        16 & 7.18 & 0.0 & 0.005 \\
        17 & 0.91 & 0.0 & 0.005 \\\hline
    \end{tabular}
\end{table}

\section{Results} \label{sec:results}

Generally, parameters inferred from large photometric surveys are analyzed from the perspective of population statistics, rather than studying individual objects.  In part, this is because the fits for any individual galaxy are often not well constrained, with star formation rates particularly poorly-determined.  This is even more relevant when the SED fitting is performed with the addition of new parameters, which increases random uncertainty and possibly causes additional degeneracies or strong covariances between parameters.  Therefore, it is necessary to consider quality cuts to maximize the statistical confidence of the fit results and identify potential outliers arising from the existing degeneracies.  The relevant selections and validation of photometric redshifts are described below.  Additionally, the subsections here show some of the well-known relationships, as well as the new ones driven by the change of the IMF shape.

\subsection{Quality Cuts} \label{sec:quality_cuts}

At first, the results of the SED fitting were used to discard poor samples.  The data here was cut based on photometric coverage to ensure that constraints on the IMF were as strong as possible.  Then, the samples where no solutions could be found or they were poorly defined were discarded.  

First, it is crucial to have a sample observed over as broad of a set of bands as possible in order to be able to constrain the solution parameters.  Thus, the redshift range was reduced to include objects in the range $0 < z < 2$, as too many photometric bands drop out at $z > 2$.

It was demonstrated in \citet{sneppen2022}, using mock SED observations, that the IMF parameter can be safely recovered only for the objects with the highest S/N ratios and at the IMF temperature above $\sim 20-25$ K.  Most of the objects in the COSMOS2020 catalog lie near the detection limit and thus have relatively low-quality flux measurements, which results in weaker constraining power for the given number of parameters and, ultimately, inferred physical properties that cannot be trusted.  For example, some solutions were often found to be erroneously fit by the most bottom-heavy or bottom-light IMFs or, alternatively, by the IMFs at edges of the sampled $T_{IMF}$ range.  As such, some faint quiescent galaxies with high mass-to-light ratio and weak constraints in the UV can be sometimes fitted with the most bottom-heavy IMF or a bottom-light IMF and high dust extinction.  

Therefore, \cite{sneppen2022} and \cite{steinhardt2022_III} defined the quality cut $S/N>10$ in the Suprime-Cam $V$ band or UltraVISTA $K$ band, at which the number of such objects best fitted at the boundaries of the temperature range dropped significantly to $\sim 10-15\%$.  Here, instead most of the objects at the temperature boundaries are discarded as having poorly-defined solutions in their chi-squared landscapes $\chi^2(T_{IMF})$ at $10 < T_{IMF} < 60$ K with steps of 1 K.  Specifically, the solutions with $\chi^2_{max}-\chi^2_{min} < 0.1$, $FWHM(T_{IMF}) < 4$ K were discarded.  If several solution temperatures were identified, the ones with the largest area $\sum_{T_{IMF}} {\chi^2(T_{IMF})}$ were selected.  As a result, $\sim 250,000$ samples out of $\sim 437,000$ (the 437,000 objects were selected as described in \S~\ref{sec:data}) passed with best-fit temperatures, out of which only 60\% of objects had $S/N < 10$.  The fraction of any $S/N$ at the margins of the temperature range did not exceed 7\%.  On the other hand, the discarded sample was the noisiest, with 90\% of its objects having $S/N < 10$ in $K_S$ flux.

By cutting objects based on the quality of their $\chi^2$ peaks, the final sample comprised 249,494 objects with 205,750 in the range $0<z<2$ where they were still covered by most of the photometric bands.  The sample was reduced even further to 156,389 objects by flagging the remaining solutions as outliers.  Those were identified as faint objects with too low or too high mass-to-light ratios and, thus, likely suffering from degenerate template parameters (more description is included in Appendix~\ref{sec:app_outliers}).  Alternatively, some objects discarded as outliers had additional local minima, similar to objects identified in \cite{sneppen2022}.  They could arise due to incorrectly fitting the highest-mass breakpoint in the Kroupa IMF in galaxies that had recently lost their most massive stars, but still have the intermediate-mass population.  Lastly, only 72,805 objects in the catalog have uncertainty in $T_{IMF}$ estimated using the minimum variance bound, as described in \S~\ref{sec:imf_temp}, as the rest of them had low significance of chi-square solutions.  These best objects were used to obtain the results in \S~\ref{sec:imf_temp}--\ref{sec:hrdiagram}.

\subsection{Stellar Mass Completeness}

\begin{figure*}[ht]
    \centering
    \includegraphics[width=0.9\textwidth]{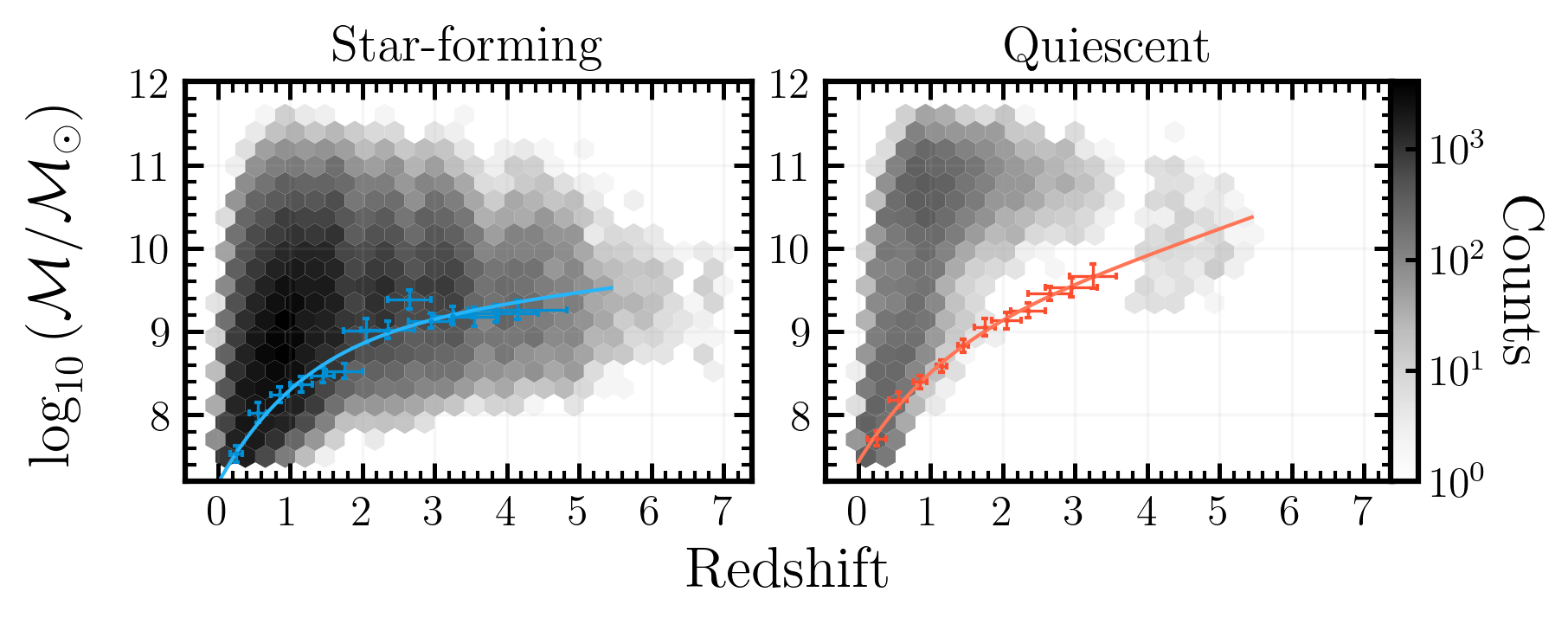}
    \caption{Counts of all galaxies in the bins of photometric redshift and stellar mass.  The points show 95\% mass completeness limits estimated by rescaling down to the detection limit in IRAC Channel 1 for star-forming and quiescent.  These limits $\log_{10}{\mathcal{M}}(z)$ were fitted with a Schechter function and shown as solid lines.  The uncertainties are taken as the median errors in bins of redshift and rescaled mass.  The levels for star-forming galaxies are below the limits of quiescent galaxies by $0.2-0.3$ dex at $0<z<2$, which reflects the higher $\mathcal{M}/L$ ratio for the passive population.}
    \label{fig:completeness}
\end{figure*}

The lower limit of stellar mass of a representative sample of galaxies depends on the survey depth and the mass-to-light ratio of galaxies and evolves with redshift.  Owing to the lack of UV and optical emission per stellar mass, quiescent galaxies are expected to be less complete than the brighter star-forming sample at every redshift in the photometry of COSMOS2020.  In addition, it is expected that at $0.0 < z < 2.0$ the galaxy sample is missing a population of blue dwarf galaxies, which fall outside of the detection image, and, possibly, some galaxies at the large stellar mass end at the lowest redshift due to a smaller volume.  The separation of galaxies into star-forming and quiescent samples was done in the UVJ color space \citep{williams2009,arnouts2013}.

By using the approach of \cite{pozzetti2010}, the limiting mass was defined by taking the faint objects that are likely to have complete stellar masses in bins of redshift and rescaling their masses to the magnitude of the survey depth.  Following the prescriptions in \cite{weaver2022b}, 1\% of the photometric fits with the largest chi-squared were removed from the sample.  Then, the stellar masses $\mathcal{M}$ of the 30\% faintest objects in IRAC Channel 1 magnitudes $m$ were rescaled to the detection limit $m^\prime = 26.4 \pm 0.1$ \citep{weaver2022a} to obtain $\mathcal{M^\prime}$:

\begin{equation}\label{eq:m_rescaled}
\begin{array}{c}
\log_{10}{ \mathcal{M^\prime} } = \log_{10}{ \mathcal{M}} + 0.4 (m - m^\prime).
\end{array}
\end{equation}

The 95\% completeness level $\log_{10}{\mathcal{M}_{95}}$ was taken as the 95th percentile of the $\log_{10} {\mathcal{M^\prime} }$ distribution and is shown with dots on Figure~\ref{fig:completeness}.  The calculation is done separately for star-forming and quiescent galaxies to account for their different $\mathcal{M}/L$ ratios.  The key assumption made here is that galaxies with $\mathcal{M^\prime}$ and $\mathcal{M}$ have the same relation to luminosity.  However, as the IMF correlations with the physical properties in \S~\ref{sec:imf_temp}--\ref{sec:props} imply, $\mathcal{M}/L$ of star-forming galaxies correlates with the stellar mass and the proper scaling should account for this difference by using $(\mathcal{M^\prime}/L^\prime)^{\beta_1}$ and $(\mathcal{M}/L)^{\beta_2}$, where $\beta$ increases for larger stellar mass (i.e. $\beta_1 > \beta_2$).  By assuming that $\beta_1 = \beta_2$, as it is done here, the rescaled stellar mass $\mathcal{M^\prime}$ and, consequently, the mass completeness are overestimated, albeit insignificantly, as the difference between the 30\% faintest galaxies in magnitude at the same redshift should be minor.  Therefore, the completeness used here acts as conservative upper limit.  Finally, although the IRAC bands were not included in the detection image of COSMOS2020, \cite{weaver2022b} argue that this should not affect the results significantly.

The completeness levels were calculated in bins of redshift at $0.1 < z < 3.4$ for quiescent galaxies and $0.1 < z < 4.3$ for star-forming galaxies in steps of 0.3.  Finally, the 95\% mass completeness levels $\mathcal{M}$ as a function of $z$ were fitted with a Schechter function (\citealp{schechter1976}; shown as solid lines in Figure~\ref{fig:completeness}; best-fit parameters are shown in Table~\ref{tab:schechter}):

\begin{equation}\label{eq:schechter}
\begin{array}{ll}
    \log_{10}{\mathcal{M}(z)} &= (0.4\ln{10}) \log_{10}{\mathcal{M}^\star} \times \\ 
    &\times[10^{0.4(z^\star-z)}]^{\alpha+1} \exp{(-10^{0.4(z^\star-z)})},
\end{array}
\end{equation}

where uncertainties were taken as median $\sigma_z$ in the bins of redshift and bins of stellar mass $\sigma_{\log_{10}{\mathcal{M}}}$ defined as $\log_{10}{\mathcal{M}_{95}} \pm 0.2 \log_{10}{\mathcal{M}}$.

\begin{table}
\centering
    \caption{Best-fit parameters and their statistical uncertainties for the Schechter function representing the limits of mass completeness for star-forming and quiescent galaxy samples (see Eq.~\ref{eq:schechter}).}
    \label{tab:schechter}
    \begin{tabular}{lccc}
    \hline\hline
    Sample       & $\log_{10}{\mathcal{M}^\star}$ & $z^\star$ & $\alpha$\\\hline
    Star-forming & 9.56$\pm$0.53 & -1.63$\pm$0.28 & -1.01$\pm$0.01\\
    Quiescent    & 9.10$\pm$0.24  & -1.89$\pm$0.13 & -1.03$\pm$0.01\\\hline
    \end{tabular}
\end{table}

This work applies two separate 95\%-completeness limits to star-forming and quiescent galaxies estimated in \cite{weaver2022b} for all relationships, except the redshift validation in Figure~\ref{fig:redshifts} and the change of properties due to the IMF modification in Figure~\ref{fig:mass_sfr}.

\subsection{Cross-check of Photometric Redshifts} \label{sec:redshifts}

\begin{figure*}
    \centering
    \includegraphics[width=0.97\textwidth]{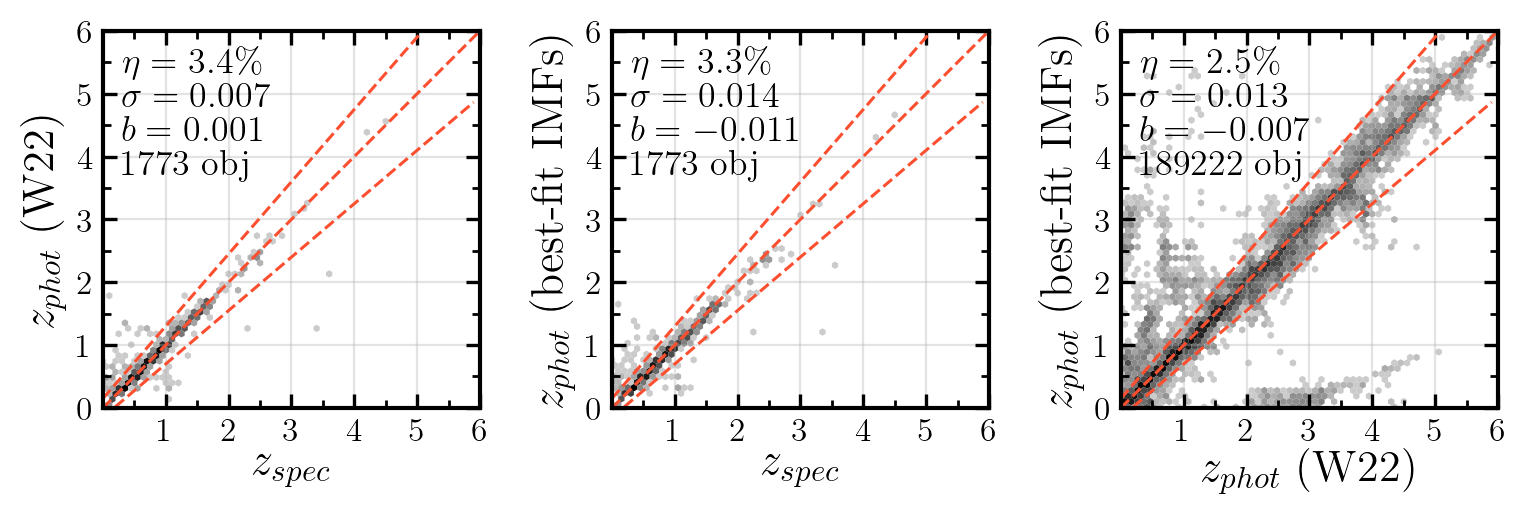}
    \caption{Density diagrams showing the validation of photometric redshifts $z_{phot}$ obtained with the best-fit IMFs.  Top panel: comparison of redshifts obtained with the Chabrier IMF at 20K \citepalias{weaver2022a} against the spectroscopic redshifts $z_{spec}$ from (Jin et al. in prep.) (1,773 objects).  Middle panel: redshifts obtained with the Kroupa IMF at the best-fit temperature against the spectroscopic sample.  Right panel: comparison of the photometric redshifts $z_{phot}$ obtained with the Kroupa IMF at the best-fit temperature and with the Chabrier IMF at 20 K (189,231 objects).  The fraction of higher than one-$\sigma$ outliers is indicated as $\eta$.  Overall, the redshifts constrained with the new SED templates appear to be consistent with the standard COSMOS2020 templates and the spectroscopic sample.}
    \label{fig:redshifts}
\end{figure*}

Spectroscopic redshifts have usually been used as a partial validation of photometric template fits.  Here, the same check is applied to the variable-IMF fits in the presented catalog.  
Figure~\ref{fig:redshifts} demonstrates the correspondence of the sample with the spectroscopic redshifts from the latest Super-deblended catalog (Jin et al., in prep.) and standard photometric redshifts based on Chabrier IMF (at 20 K in the Milky Way) from the COSMOS2020 survey \citep{weaver2022a}.  The outliers in the one-to-one comparison are defined as points satisfying the condition $\vert \Delta z \vert > 0.15(1 + z)$ \citep{hildebrandt2009}.  The fraction of outliers $\eta$ is indicated on the plot panels.  It shows that the spectroscopic sample disagreed in $\sim 3$ \% of cases, similar to the number outliers in the standard catalog, as shown in Figure~\ref{fig:redshifts} and in \cite{weaver2022a}.  The precision of the photo-$z$ is estimated by using the normalized median absolute deviation $\sigma_{NMAD}$ \citep{brammer2008}, which is not strongly sensitive to outliers: 

\begin{equation}
\begin{array}{c}
     \sigma_{NMAD} = 1.48 \times {\rm median}\left( \frac{\vert \Delta z - {\rm median}(\Delta z) \vert}{1+z_{spec}} \right)
\end{array}
\end{equation}. 

Finally, the bias in the estimates is denoted as $b={\rm median}(z_{phot} - z_{spec})$.  These metrics perform similarly to the standard measurements reported in the COSMOS2020 catalog.  The agreement confirms that even more significant differences in templates have little effect on the redshifts reconstructed with the same code \citep{sneppen2022}.  Finally, the outliers identified in Figure~\ref{fig:redshifts}, have been discarded from the scientific results in the rest of the work.

\subsection{IMF Temperatures} \label{sec:imf_temp}

\begin{figure*}
    \centering
    \includegraphics[width=\textwidth]{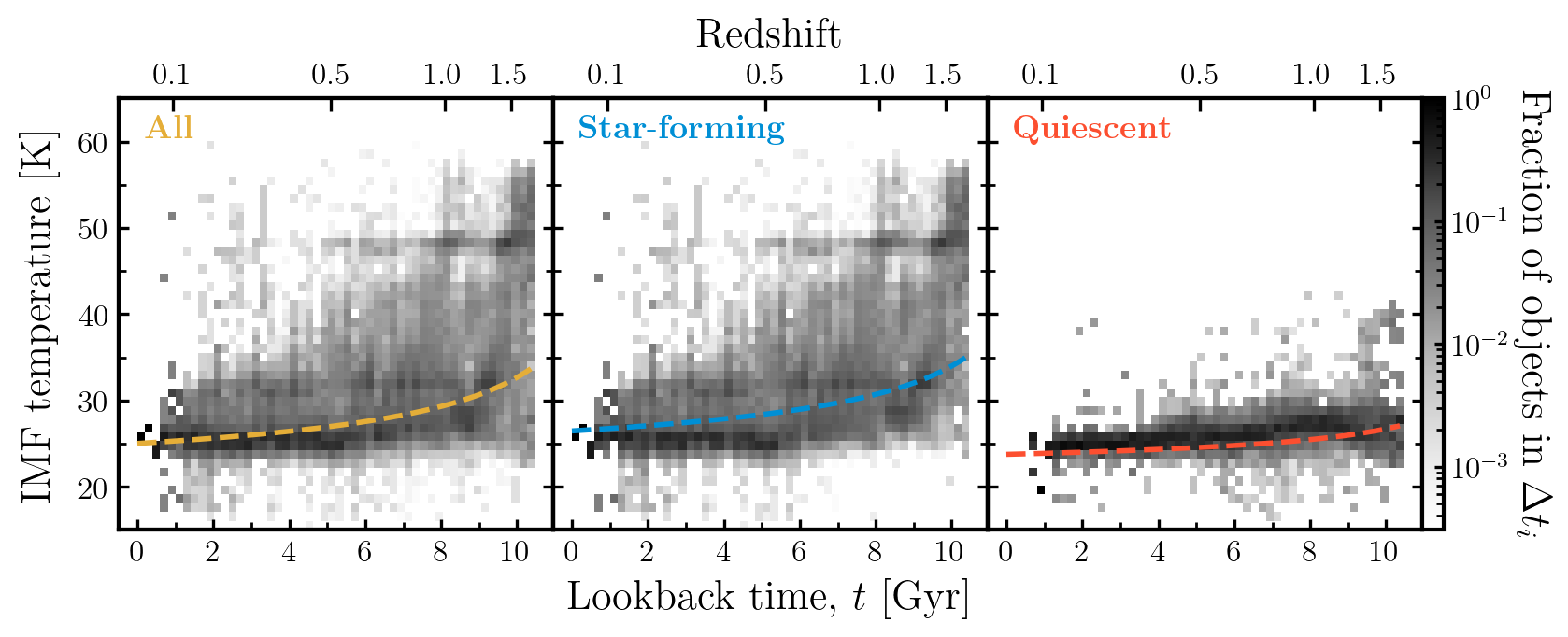}
    \caption{The number of objects in bins of IMF temperature and lookback time, normalized in lookback-time bins $\Delta t_i$, for the total (left), star-forming (middle) and quiescent sample (right).  Dashed lines show best-fit exponentials fit in the space of temperature and redshift $(T, z)$ using individual non-normalized data.  Uncertainties in redshift are based on $1\sigma$ width of $p(z)$ solutions from \texttt{EAZY} (median binned errors range between $\sigma_z \sim 0.01-0.27$).  The errors in IMF temperature are computed using the likelihood based on chi-squared of the SED solutions, and are very loose (see text for details; median binned errors range between $\sigma_{T_{IMF}} \sim 3-12$ K).  Most of the objects have IMF temperature above that of the analogs in the local universe forming a continuum of bottom-lighter IMFs.  On average, the best-fit IMF temperature increases with redshift as in \cite{sneppen2022,steinhardt2022_II}.  Finally, the temperature and corresponding IMF behaviour is distinct between active and passive galaxy populations, where the former is almost exclusively responsible for the temperature increase in the total sample.  This result for star-forming and quiescent galaxies is consistent with the observed trends in temperature of cold dust in similar redshift ranges (see text for references).}
    \label{fig:temps}
\end{figure*}

The main result in this work is that most of the galaxies are unlike the local analogs and have bottom-lighter IMFs, as traced by the IMF temperature parameter (Figure~\ref{fig:temps}).  These best-fit IMFs indicate the larger relative abundance of the most massive stellar populations at all redshifts.  Theoretically, this change is expected to correspond to hotter (and, presumably, denser and metal-poorer) molecular gas.

Possible relationships in the distribution of IMF temperatures $T_{IMF}$ at different redshifts $z$ are investigated by fitting an exponential function $T_{IMF}(z) = A \exp{(-z/\tau)}$ (dashed lines in Figure~\ref{fig:temps}).  There is not a proper way in the SED-fitting procedure implemented here for estimating uncertainties in temperature, so the errors are calculated using the chi-square $\chi^2$ statistic computed by \texttt{EAZY} for best-fit models produced at each IMF temperature in $10 < T_{IMF} < 60$ K in steps of 1 K.  By assuming that the temperature estimator is unbiased (at $T_{IMF} \gtrsim 20-25$ K) and using the Minimum Variance Bound, the uncertainty $\hat{\sigma_{\hat{T}}}$ was calculated as $\Delta \hat{T}_-$ and $\Delta \hat{T}_+$ at $\log{L(\hat{T} \pm \hat{\sigma}_{\hat{T}})} = \log{L_{max}} - 1/2$ with the likelihood as $\log{L} = -2 \chi^2$.  For the solutions that did not have the sufficient significance, the uncertainty fields in the catalog are left empty.  The best-fit function in the total sample (left panel) shows that the temperature increases from $\sim25$ K to 35 K at $0 < z < 2$, mostly driven by star-forming galaxies which were selected in UVJ color space \citep{williams2009,arnouts2013}.

Although there is a large scatter in solutions for IMF temperature in the whole sample, there is a systematic trend for increasing temperature at higher redshift.  One component of the scatter is added by the fitting procedure and is expected to be mostly random at temperatures above $\sim 25$ K.  Based on the fits of synthetic photometry, \cite{sneppen2022} showed that scatter in the recovered temperature is $\sigma \approx 5-10$ K and the estimator is only slightly biased at those temperatures for the strongest photometric detections, while almost completely inaccurate for ``cooler'' galaxies (see Figure 13 in their publication).  However, it is possible that other unaccounted systematic effects contribute to the obtained temperature distribution.  For example, one of the critical limitations of the SED templates employed here is that the temperature, density and metallicity of the star-forming gas are not causally connected to the expected changes in the definition (Eq.~\ref{eq:imf}) of the IMF.  Therefore, the resulting ``temperature'' quantity is likely a combination of effects due to these three quantities.

The rest of the scatter appears to be due to the differences in the mass to light ratio, $\mathcal{M}/L$, of stellar populations in different galaxies.  For example, the most common and strong distinction exists between active and quiescent galaxies, where the latter are expected to have bottom-heavier IMFs.  Indeed, the right panel in  Figure~\ref{fig:temps} shows that the passive population is consistent with the same $\mathcal{M}/L$ and has relatively small scatter.  Given that the quiescent galaxies with very weak UV emission are expected to be the least sensitive to probing the IMF, this result validates the fitting procedure.  This would be consistent with the measurements of dust temperatures in quiescent galaxies that appear to have a lower limit at $T_{dust} \sim 21 \pm 2$ K in a similar redshift range in \cite{magdis2021}.  However, it is possible that the temperatures of the star-forming gas and dust are set by different mechanisms in quiescent galaxies and therefore the former does not have to be similarly bounded at $\sim 20$ K.  This possibility cannot be tested with the IMF estimator here, as it is completely biased in that temperature regime.  On the other hand, the active population spans the full sampled temperature range at every epoch and is almost entirely consistent with the temperature increase in the total sample.

By isolating the quiescent galaxies, it appears that the large scatter is indicative of various star-forming galaxies.  While most appear to be on the main sequence the rest can represent different sub-populations.  Among others, ULIRGS that are thought to result from major mergers of gas-rich galaxies or close interactions (from observations \citealp{sanders1991,solomon1997}; from simulations \citealp{mihos1996}), have been shown to exhibit a constant IMF temperature at $\sim 30$ K up to $z=2$ \citep{sneppen2022}, consistently with dust temperature estimates in \cite{bethermin2015}.  Among galaxies efficiently forming new stars are ``green peas'' \citep{cardamone2009} that have small stellar mass and compact morphology, little dust obscuration, and other starbursts that include galaxies with varying physical properties but appear to have similarly high sSFR due to a recently ignited burst of star formation \citep{kennicutt2012}.  Based on their properties, these types are likely to occupy the higher temperature tail, above ULIRGS and typical SFMS galaxies.  Indeed, \cite{zhang2018} showed that starbursts at $2<z<3$ have a top-heavy IMF by measuring $^{13}{\rm C}/^{18}{\rm O}$ transitions.  Note, the density in Figure~\ref{fig:temps} reveals a grid pattern, which is likely due to fits with degenerate parameters, described in Appendix~\ref{sec:app_outliers}, that could not be removed entirely.

Finally, the large scatter in the IMF shapes here is in agreement with the broad distribution of equivalent widths of H$\alpha$ at $z \sim 2$ in \cite{nanayakkara2017}.  That work shows that the most likely contribution to the large widths is the top-heavy IMF.  Moreover, the large scatter in the widths cannot be explained entirely by the starbursts and likely indicates either stochasticity or some underlying relationships in the shapes of the IMF.

This behavior of the IMF shape is similar to that found using galaxies in COSMOS2015 \citep{sneppen2022}, although it now extends to fainter galaxies and lower temperatures at the higher redshift and includes galaxies with higher measurement quality on average.  In particular, more data fills the lookback time $t > 8$ Gyr ($z > 1.1$) and $T_{IMF} < 30$ K in Figure~\ref{fig:temps}.  It represents a faint red sample at the lower tail of sSFR or faint blue galaxies, which have been observed owing to the depth of the COSMOS2020 surveys.

\subsection{Physical Properties of Galaxies} \label{sec:props}

\begin{figure*}
    \centering
    \includegraphics[width=0.75\textwidth]{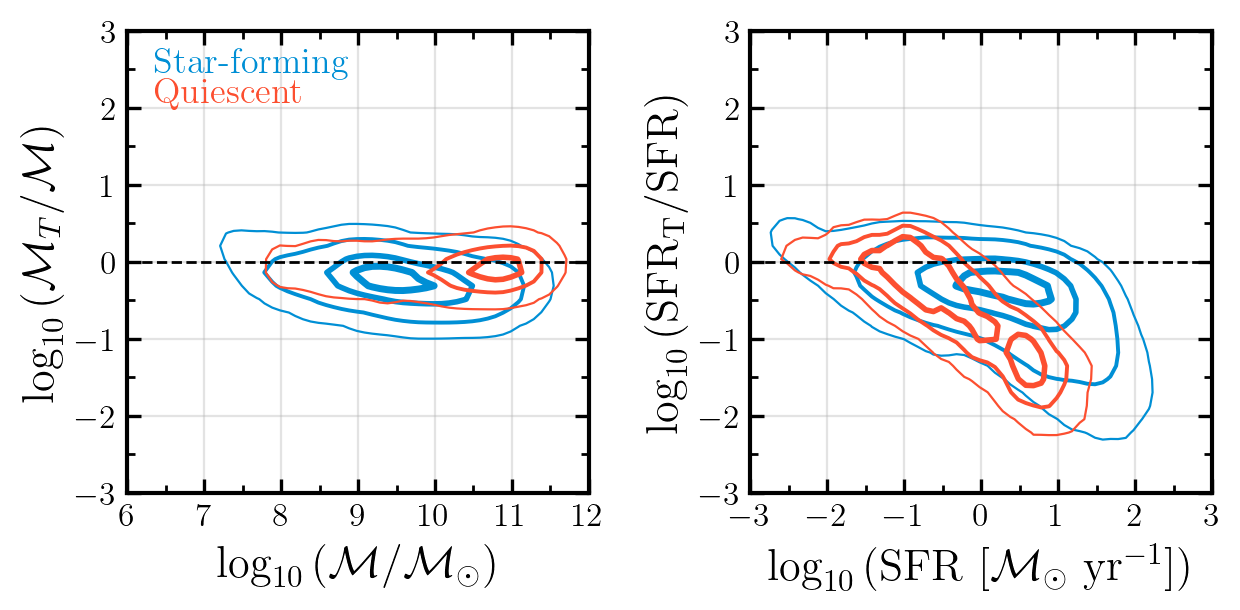}
    \caption{Difference in stellar mass and star formation rate between the ``best-fit IMF sample'' $(\mathcal{M}_{T}, {\rm SFR}_{T})$ and the standard properties from COSMOS2020 $(\mathcal{M}, {\rm SFR})$ at $0 < z < 2$.  The properties $(\mathcal{M}_{T}, {\rm SFR}_{T})$ are taken at the best-fit IMF temperature $T$, while $(\mathcal{M}, {\rm SFR})$ are based on the local IMF, analogous to $T \sim 20$ K.  The standard COSMOS2020 properties were converted from Chabrier to Kroupa IMF for a consistent comparison, scaling properties by 1.06 (eg., $\mathcal{M_K} = 1.06 \mathcal{M_C}$, using the conversion from \citealp{zahid2012}).  The separation of galaxies into star-forming and quiescent was done in the $UVJ$ color space of the best-fit IMF colors.  The $UVJ$ classification is consistent between two catalogs with 98.2\% agreement.  The contours outline 42,545 active and 8,416 passive galaxies.  Clearly, as the best-fit IMF is almost always bottom-lighter than in the Milky Way, the masses and star formation rates shift to lower values.  The median differences of the total sample correspond to $log(\mathcal{M}_{T}/\mathcal{M}) \approx -0.21$ (1/1.6 linear change) and $log(SFR_{T}/SFR) \approx -0.40$ (1/2.5 linear change).  The offset is on average higher for star-forming galaxies with the highest SFR, while quiescent galaxies experience little to no change in stellar mass and a similar change in SFR.  The contours in both plots are consistent and extend to the same number levels (outer contours cover 98\% of the respective total numbers).}
    \label{fig:mass_sfr}
\end{figure*}

If most studied galaxies have systematically different stellar populations from what was assumed previously, their inferred stellar mass and the rate of star formation must change correspondingly.  Figure~\ref{fig:mass_sfr} shows the changes in these two properties at $0.2 < z < 2.0$.

Clearly, most of the objects have properties characteristic of the bottom-lighter IMFs, or higher temperature of the star-forming gas.  As can be expected, the bias is associated predominantly with the bottom-lighter IMF solutions for the star-forming galaxies.  There, the magnitude of the change appears to be correlated with the SFR of galaxies, or the strength of UV emission.  The largest offset reaches $\sim1$ dex in stellar mass and $\sim 2$ dex in SFR, while the median changes are $\sim 0.2$ and $\sim 0.4$ dex, respectively.  

The best-fit properties for the quiescent population at these redshifts do not appear to be affected significantly by the change.  This indicates two possibilities that: (1) their last stellar populations formed at the same temperature as in the local universe; (2) photometric IMF variations in the quiescent galaxies are below the sensitivity limit of our procedure of $T \sim 20$ K.  Nevertheless, the most bottom-light quiescent galaxies are still found in the lowest redshift bins, at $z < 0.8$, which suggests that some internal feedback (stellar radiation or cosmic rays) on top of the CMB may heat the gas at the time of the last star-forming episode to above 25 K at $z > 0.8$.  However, it is also likely that some of coldest faint quiescent galaxies at $z>0.2$ may still be undetected.

It is also worth noting that the photometric fitting method is not capable of recovering IMF shapes with IMF temperature $T < 20$ K, as show in the mock simulations in \cite{sneppen2022}.  Therefore, the lower bound found here is characteristic of the method.  Although physically this bound coincides with the temperature background in the MW, it disagrees with the probes of even bottom-heavier IMFs in elliptical galaxies and other old stellar structures at low redshift in \cite{conroy_vandokkum2012,vandokkum_conroy2012,geha2013,martinnavarro2015,conroy_vandokkum2017,vandokkum_conroy2017,hallakoun2021}.

\subsection{Star-forming Main Sequence} \label{sec:sfms}

\begin{figure*}
    \centering
    \includegraphics[width=0.95\textwidth]{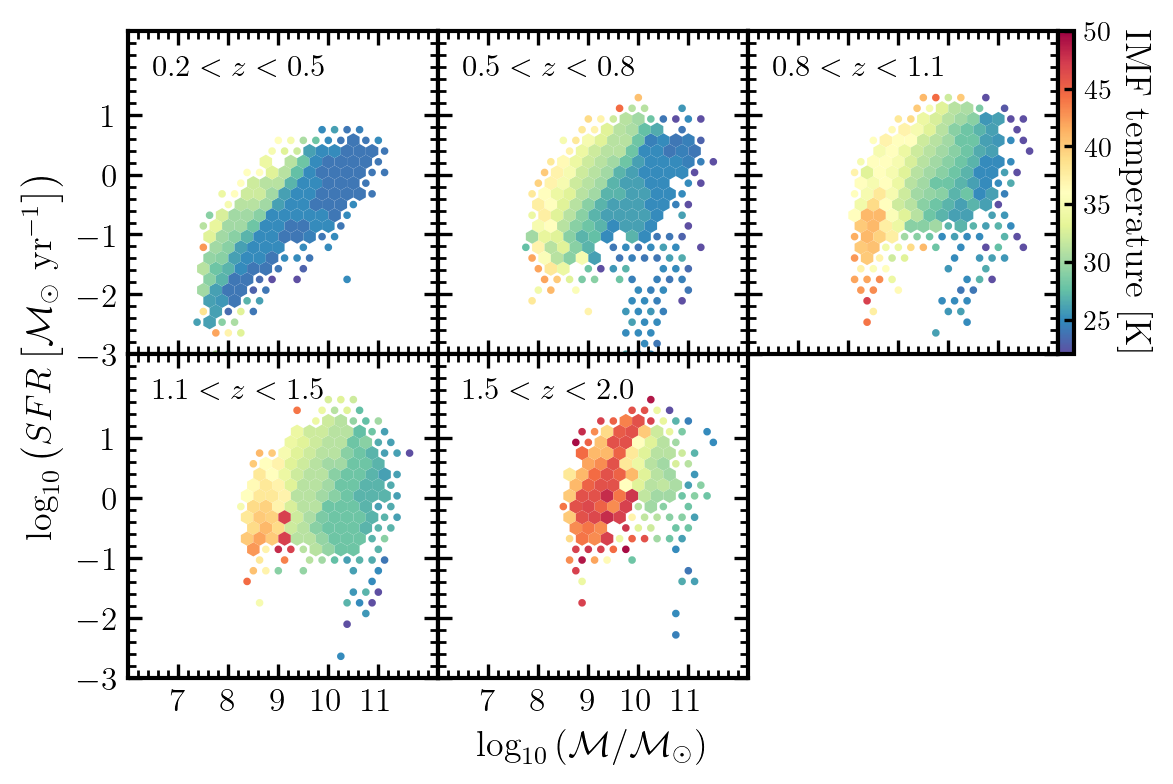}
    \caption{All galaxies plotted in the space of the star-forming main sequence in redshift windows from $z=0.2$ to 2.0.  The color corresponds to the median IMF temperature in a bin of stellar mass and SFR ($log(\mathcal{M})$, $log(SFR)$).  Bins with $\geq 10$ samples are filled, bins with $1-9$ samples are shown as scatter.  Typical star-forming galaxies are found on the main sequence, showing a gradient of IMF temperatures correlated with the sSFR; quiescent galaxies turn off the sequence and extend to lower SFR at the highest stellar masses and lowest IMF temperatures; objects with the highest IMF temperatures at the low stellar mass end exhibit a distinct temperature relationship from the typical main sequence (the temperature relationship is shown on Figure~\ref{fig:hr_diagram_20_15}).}
    \label{fig:sfms}
\end{figure*}

At every redshift, most of the actively star-forming galaxies of each stellar mass have been found to exhibit a narrow range of SFR, which defines the star-forming main-sequence (SFMS) \citep{brinchmann2004,noeske2007,peng2010,speagle2014}.  This relationship holds in both photometrically and spectroscopically-derived SFR and despite the fact that galaxies have a range of different star formation histories, supernova rates, strengths of active galactic nuclei feedback or cosmic environment.  However, it can still be expected that variability in these factors can play role in changing the distribution of stellar masses in a galaxy.  Therefore, one of the ways of testing the properties of the variable IMF fits is by verifying that the main sequence still holds.

Figure~\ref{fig:sfms} shows all galaxies, where the star-forming ones still form the SFMS at each redshift despite the shift in stellar masses and star formation rates due to changing the Chabrier IMF at 20 K to IMFs at the best-fit temperatures.  Similarly, the work of \cite{steinhardt2022_II} that implemented the same temperature-dependent IMF, showed that there is a correlation between SFR, stellar mass at every redshift, consistent with the known relationship.  While the SFMS appears to be the same qualitatively, the exact offset and gradient of the relation should be sensitive to the shape of the best-fit IMF and should correlate with it, as suggested by the change of the SFR and stellar mass in Figure~\ref{fig:mass_sfr}.  

The best-fit properties here indicate that there is a gradient of IMFs along the SFMS at every redshift.  The IMF temperature increases most rapidly towards the high SFR and low stellar mass end.  The gradient becomes steeper at the higher redshift, with more extreme cases of the high-temperature galaxies.  As shown in Figure 1 of \cite{steinhardt2023}, the change of IMF shape may be correlated with the gradient of the temperature of dust continuum emission shown in \cite{magnelli2014}.  This suggests that the ``IMF temperature'' parameter may trace the temperature of molecular clouds at star formation, as argued theoretically \citep{jermyn2018}.  Therefore, with the evidence presented here and in the previous work it may be compelling to model a possible relationship or an equilibrium state between temperature of the molecular clouds and cold dust.  If calibrated at low redshift, this relationship can be used to constrain the IMF for galaxies at high redshift for which measuring gas temperatures from line ratios is challenging (see \S~\ref{sec:imf_prior} for further discussion).

As expected, the quiescent population is found at the highest mass bins for a given SFR and redshift.  The colors in Figure~\ref{fig:sfms} indicate that the quiescent galaxies are coolest at the lowest redshift, and gradually become warmer at the higher redshift.  This effect demonstrates that either the star-forming gas used to form the last stellar population is hotter in earlier epochs due to some feedback mechanisms or that some of the coolest samples at the higher redshift are below the detection limit of the COSMOS2020 surveys.  The last is more likely, as observations show that the average temperature in massive quiescent galaxies at $0.2 < z < 2.0$ is approximately constant at $T_d = 21 \pm 2$ K \citep{magdis2021}.  Finally, the quiescent galaxies are coolest among other galaxies at every epoch.  Thus, they appear to turn off the main sequence once some critical mass is built up\footnote{Evidence in \S~\ref{sec:smf} from quiescent galaxy fraction appears consistent with there being a characteristic mass for quenching at every redshift, which monotonically decreases towards the current epoch.} which appears to be traced by the IMF temperature.

A population of low-mass and hot galaxies appears as indicated by the IMF temperature in the highest redshift panels of the SFMS.  These galaxies have distinct combination of sSFR and IMF temperature, as discussed in more detail in \S~\ref{sec:hrdiagram}.  \cite{steinhardt2023} suggested that they also appear to have compact morphologies.  It has been hypothesized that these galaxies may represent the early, core-forming stage of galaxy evolution.  If that is the case, this population is only cooling down onto the SFMS and thus can be expected to have a different $SFR-M_\star$ relationship or none, similarly to quiescent galaxies.  

It is unclear whether this sample is separate from the typical star-forming galaxies on the main sequence and therefore requires investigation.  Previously, it has been argued by \cite{abramson2014} that the SFMS can be explained solely by the spatially-resolved evolution of the star-forming disks which host evolved bulges.  If the fundamental SFMS evolution is driven purely by the physics of star formation, it is possible that the core-forming galaxies form a similarly straight SFMS.  On the other hand, if the typical main sequence galaxies are driven by a combination of star formation and disk-morphology effects, the core-forming galaxies may exhibit a characteristically different main sequence: a power law with a different exponent, normalization or a different functional form (more details follow on this in \S~\ref{sec:hrdiagram}).  Thus, morphology may present hints for the mechanisms driving the SFMS relation and other stages of evolution.

Finally, the galaxy sample covers almost 4 orders of magnitude in SFR with some of the fainter samples detected only in COSMOS2020 and not the previous COSMOS catalogs (Figure~\ref{fig:sfms}).  The low-mass galaxies are cut sharply at the higher redshift due to a combination of the Malmquist bias at the low SFR, the quality cuts here at the higher SFR (see \S~\ref{sec:quality_cuts}) and mass completeness cuts.  At the same time, the most massive and luminous galaxies are found in the most distant observations where the sampled cosmic volume is largest.

\subsection{Stellar Mass Function} \label{sec:smf}

\begin{figure*}
    \centering
    \includegraphics[width=0.8\textwidth]{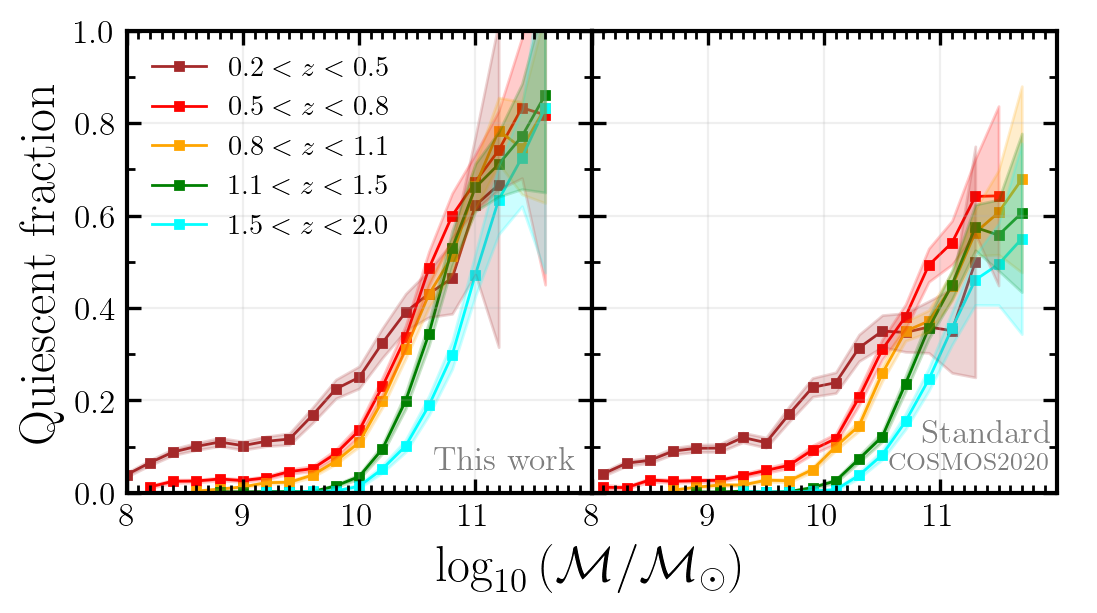}
    \caption{The observed fraction of the quiescent galaxies as a function of stellar mass defined as $log( \Psi_{QS} / [\Psi_{SF} + \Psi_{QS}])$ for a stellar masses derived with a best-fit IMF (left) and standard stellar masses form the COSMOS2020 catalog (right).  The uncertainties include Poisson error in star-forming and quiescent galaxy counts.  Stellar mass $log(\mathcal{M})$ is limited according to completeness levels at different redshift for star-forming and quiescent galaxies \citep{weaver2022b}.  At every redshift, most massive galaxies complete their evolution first, and lowest mass galaxies take the longest time to evolve.  The evolution from high to low mass end progresses smoothly with redshift.  Overall, this picture is consistent with the effect of downsizing.  Notably, for the properties with the best-fit IMF this effect appears consistently at all redshift bins -- even at $z > 1$ -- unlike in some of the previous studies of stellar mass functions (see text for details).  These properties (left) show a systematically higher fraction of quiescent galaxies in the most massive tail than inferred from the standard property set (right).  This difference comes from the differential change in the stellar mass, where the star-forming functions are shifted towards lower masses, while the quiescent ones mostly remain unchanged (this comparison is shown in Appendix~\ref{sec:appendix_smf}).}
    \label{fig:smf}
\end{figure*}

Stellar mass function, as the comoving volume density of star-forming and quiescent populations at different redshift snapshots, is an observational tracer of growth of stellar mass at different scales over time \citep{fontana2006,drory2008,marchesini2009,peng2010,pozzetti2010,ilbert2013,moustakas2013,muzzin2013,grazian2015,song2016,davidzon2017,stefanon2021,weaver2022b}.  These observations provide constraints for theoretical models and simulations of galaxy evolution, and eventually help to study the mechanisms regulating galaxy assembly, growth and quenching of star formation.  Here, the fraction of quiescent to total galaxies in terms of the mass functions is used to show the growth and the hierarchy of the quenched population.  

The fraction of quiescent galaxies as a function of stellar mass obtained with the best-fit IMFs is reproduced with the same overall characteristic behaviour as in most studies that observe the effect of downsizing, reported first in \citep{cowie1996,juneau2005,fontanot2009}.  The mass scale of quenching on Figure~\ref{fig:smf} indicates that the most massive galaxies are formed at earlier epochs and evolve faster than the less massive populations at lower redshift.

According to the best-fit stellar masses obtained with a bottom-lighter variable IMF (Figure~\ref{fig:mass_sfr}), the stellar mass function for star-forming galaxies is shifted towards the lower stellar mass with respect to the quiescent which stays unaffected (individual mass functions are shown in Appendix~\ref{sec:appendix_smf}).  Therefore, the passive galaxy population becomes more abundant for the most massive galaxies at every redshift.  As a result, the evolution of the quiescent mass scale is even more sharply consistent with mass downsizing in comparison to standard photometric properties, similarly to results in \cite{steinhardt2022_III}.
   
The change in the quiescent galaxy fraction with redshift allows placing constraints on possible mechanisms of quenching.  As shown in \cite{steinhardt2022_III}, there appears to be a characteristic mass at every redshift, above which majority of the galaxies have stopped forming stars.  This characteristic mass monotonically increases with redshift, consistent with downsizing.  This behaviour is consistent with running out of cold molecular gas or some universal feedback mechanism and ability to cool it or, more broadly, with the framework of ``mass quenching'' \citep{peng2010}.  Some possible quenching mechanisms include morphological quenching, AGN feedback or cosmological starvation \citep{man2018}.

\subsection{Galaxy Population Diagram} \label{sec:hrdiagram}

This section shows a relation between the sSFR, or inverse galaxy growth timescale, and the temperature of the star-forming gas derived from the IMF constraints.  It appears that main-sequence galaxies experience a decline in the efficiency of converting their gas into stars towards lower redshift from molecular gas studies \citep{magdis2021} or, more generally, from abundance matching \citep{finkelstein2015}.  Similarly, there is a monotonic decrease in sSFR of main-sequence galaxies from photometric properties \citep{peng2010,speagle2014}.  As seen from the SFMS in this work (Figure~\ref{fig:sfms}), the IMF temperature is proportional to sSFR for star-forming galaxies, but the relationship is likely different or breaks completely for quiescent galaxies.  Should this behaviour break or change characteristically for any galaxies in our sample, this may signify a different regime of star formation, as argued in \cite{steinhardt2023}.

\begin{figure*}
    \centering
    \includegraphics[width=1\textwidth]{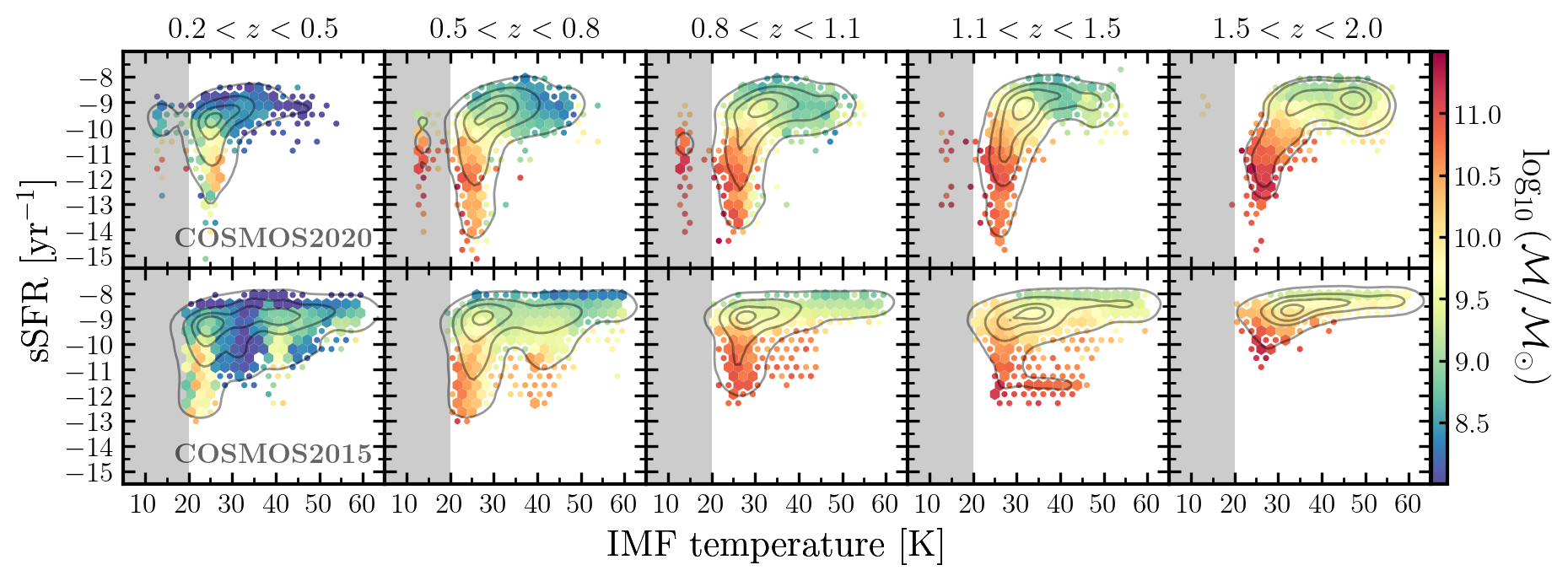}
    \caption{Diagram of sSFR as a function of IMF temperature $T_{IMF}$ for several redshift windows in the COSMOS2020 (top row; provided here) and COSMOS2015 surveys (bottom; from \citealp{sneppen2022}).  The bins are colored by median stellar mass $log(\mathcal{M})$.  The black contours show the sample density for reference.  Bins with $\geq 10$ samples are filled, while the bins with $1-9$ samples are shown as scatter.  The gray-shaded area shows the range where the temperatures cannot be recovered.  The densest regions are populated with the typical main-sequence galaxies, which appear to be driven by the coupling of sSFR and IMF temperature.  The two tails with characteristically different $sSFR-T_{IMF}$ relationships from the typical main sequence appear to be quiescent (low $T_{IMF}$) and active (high $T_{IMF}$).  These distinct galaxy regions show a smooth gradient of stellar mass, which together with the declining sSFR and $T_{IMF}$ suggests a clear evolutionary sequence.  The sample shown at the top excludes outliers resulting from spurious SED fits (see Appendix~\ref{sec:app_outliers}).  According to the synthetic tests performed in \cite{sneppen2022}, the best-fit IMF properties cannot be trusted at $T_{IMF} \lesssim 20$ K.  Therefore, sub-samples at $T_{IMF} \approx 10-20$ K that are present in most panels here cannot be trusted.}
    \label{fig:hr_diagram_20_15}
\end{figure*}

Figure~\ref{fig:hr_diagram_20_15} shows the relationship between sSFR and the IMF temperature in the sample at $0.2 < z < 2.0$.  As can be expected, the densest region of the diagram is populated by galaxies on the main sequence.  There IMF temperature and sSFR appear to be coupled the strongest for these galaxies, albeit spanning a small range of both properties.  Likely, this is consistent with the strong feedback mechanisms driving these galaxies on the SFMS.

The $sSFR-T_{IMF}$ coupling appears to break in two different ways for the extreme tails in temperature and mass, which suggests characteristically distinct galaxy populations.  The low temperature end with the most massive objects is represented by the quiescent population.  Finally, the population at the high temperature has the shortest growth timescales, lowest masses and thus is likely to be in the state of rapid star formation and cooling onto the ``typical'' main sequence.  Perhaps these properties signify three distinct modes of star formation, connected by galaxy evolution tracks first from early rapid formation, then to the strongly-regulated main sequence and, finally, quiescence.  Possible physical processes defining these modes are discussed in more detail in \cite{steinhardt2023}.

In addition, \cite{steinhardt2023} suggest that the thermal evolution on and off the star-forming main sequence may be independently linked to the apparent morphological development of galaxies.  The connection shows that the cooling off the main sequence galaxies is possibly associated with the transition from disk-like structures with a central bulge or a core to a more extended elliptical structure.  The evolution is also apparent in the change of photometric colors, from predominantly blue disks with red cores to red ellipticals.  Finally, the new, early stage of evolution appears to have blue compact cores as their morphological counterparts at all redshifts, which could be progenitors of the future disk-like galaxies with a predominatly quiescent core, similarly to the morphologically-driven evolution in \cite{abramson2014}.

Notably, $sSFR = 10^{-8}$ yr$^{-1}$ is a common upper limit set by star formation rate in SED templates in different fitting codes.  Greater sSFR are not allowed to avoid degeneracy between the star formation history and dust in solutions.  Therefore, the real gradient at the high temperature tail may be steeper than observed here.  The true slope is going to constrain the potential physical mechanism driving the evolution of this galaxy population.

On the other end, the quiescent population appears to be limited from the bottom at around $T_{IMF} \approx 20 K$, which is either real or limited by the sensitivity of the photometric procedure to the weak UV signal.  The 20 K limit is in agreement with the local molecular gas measurements in the Milky Way \citep{schnee2008}.  The Galaxy is likely on the way down in the quiescent branch at low temperature and at $sSFR \sim 10^{-11}$ yr$^{-1}$ \citep{licquia2015} on Figure~\ref{fig:hr_diagram_20_15}.  Therefore, this suggests that the sharp cut at the low temperature and at least down to the MW sSFR is real.  Nevertheless, the technique employed here cannot recover temperatures lower than $20-25$ K, as shown using mock galaxies in \cite{sneppen2022}.  Therefore, the method here is not capable of recovering those IMF that have bottom-heavier shapes than in the MW, as reported for stellar systems with old stars in \cite{conroy_vandokkum2012,vandokkum_conroy2012,geha2013,martinnavarro2015,conroy_vandokkum2017,vandokkum_conroy2017,hallakoun2021}.

\section{Discussion} \label{sec:discussion}

The analysis here shows that the new properties show a clear picture of mass-dependent assembly of galaxies in terms of the stellar mass functions and provide new valuable insights into the star-forming processes along and ``around'' the star-forming main sequence.  It is shown that the gas temperature derived from the best-fit IMF adds a new dimension to investigate efficiency of star formation and possibly dissect galaxy evolution into different stages.  This is supported by the recent work that finds a connection between IMF-selected galaxies and their morphologies \citep{steinhardt2023} and suggests a possible interpretation in which galaxies evolve from blue compact cores, through disk formation around the evolved core to elliptical extended structures.

This section emphasizes a possible bias introduced by assuming a universal IMF based on the Milky Way, as well as the recently published issues with even averaging observations of different parts of individual galaxies.  It finally discusses implications for the problem of the ``impossibly early galaxies'', which becomes relevant here by assuming that the observed increase in the temperature of the various parts of the ISM continues to high redshift, so that the bias in the estimated mass of the stellar populations becomes increasingly large.

\subsection{The Milky Way Bias} \label{sec:mw_bias}

Resolved observations in the local universe uncover the chemical and structural complexity of galaxies that translates into variations in their star formation processes \citep{hodge1989,gallart1996}.  It has been found that the cold ISM exists in a range of phases that are sensitive to the internal processes as well as the environment \citep{saintonge2022}.  The latter review shows that there are systematic variations in the amount of molecular gas and the rate at which it is converted into stars across individual galaxy populations.  These differences in the cold phase of the ISM are expected to produce different stellar populations resulting in different feedback, which can build up to produce strong differences in stellar masses, metallicities and other quantities over time.  As a result, it is easy to introduce systematic bias in inferred astrophysics by making strong assumptions that neglect these differences.  

Historically, the complexity in star-forming conditions has been dealt with by assuming that the IMF derived in the MW is universal (eg., \citealp{kroupa2001,franceschini2006,arnouts2013}), although some hints of IMF variability have become available \citep{treu2010,martinnavarro2015,vanDokkum2017}.  
However, galaxy evolution over cosmic time has been shown to break even under different IMF measurements made only in the Milky Way \citep{speagle2014}.

Evidence from cold dust continuum emission and atomic gas measurements \citep{valentino2020} suggests that molecular gas temperatures at high redshift are unlikely to be the same as in the local universe.  The dust temperature from stacked measurements increases from 20 K at $z=0$, equivalent of the MW gas temperature, to 40 K at $z=2.3$ \citep{magnelli2014} and to 50 K at $z=4$ \citep{schreiber2018,cortzen2020}.  This is consistent with the evidence of increasing intensity of the radiation field \citep{bethermin2015,magdis2017} in earlier galaxies.  Based on theoretical work, temperature changes in the star-forming gas should be sensitive for the mass scale of the stellar population \citep{jeans1902,low1976,larson1998,larson2005} and modify the shape of the IMF \citep{jermyn2018}.

Therefore, it is necessary to relax the assumption of the local IMF at least for non-local galaxies to reflect the expected changes in the star-forming conditions.  Previous studies of H$\alpha$ equivalent width (\citealp{gunawardhana2011,nanayakkara2017}; and references therein) indicated that galaxies at $z \sim 0.35$ and $z \sim 2$ have flatter high-mass IMF slopes, which likely evolve with redshift.  Besides, \cite{zhang2018} find a more direct indication of a top-heavy IMF in starburst galaxies at $2<z<3$ from $^{13}{\rm C}/^{18}{\rm O}$ lines.  Based on the photometric constraints obtained in this work and previously in \cite{sneppen2022}, the IMF at $0<z<2$ is systematically different in star-forming galaxies and is likely to change as the CMB temperature increases at higher redshift and due to a complex interplay of various feedback processes.  Consequently, the stellar masses of typical star-forming galaxies appear overestimated by up to 1.0 dex in the most extreme cases (Figure~\ref{fig:mass_sfr}).  As a result, the interpretation of such key processes in galaxy evolution as quenching becomes more pronounced - it is more consistent with the effect of downsizing (see \S~\ref{sec:smf}).  In addition, the IMF-derived temperature of star-forming gas hints at an early and hot stage of galaxy evolution \citep{steinhardt2023}, as well as provides insights into the star-forming main sequence (see \S~\ref{sec:hrdiagram}).

When it is not possible to constrain an IMF, which is the case for most current observations outside the COSMOS field, it is instead necessary to assert a physically motivated IMF, because using the 20 K local IMF can break the inferred galaxy properties in the different ways addressed in this work in \S~\ref{sec:props}--\ref{sec:hrdiagram}.  The assumption should be based on the CMB temperature at the relevant redshift regime\footnote{It is worth noting that the IMF-derived temperature theoretically corresponds to the temperature in the star-forming clouds of the photometrically-constrained stellar population.  Thus, for the quiescent population this temperature is offset from the observation by the time when the last stellar population formed.} and the latest understanding of the ISM temperature.  While there is no complete theory of the feedback-driven gas temperature in galaxies, it is possible that the temperature measurements of dust continuum or [C {\sc i}] gas line ratio in specific galaxies or galaxy populations can be used as a proxy (more on this strategy in \S~\ref{sec:imf_prior}).

Finally, the observations of galaxy populations throughout the cosmic time suggest that even if there is a most-likely universal IMF, it likely does not look like the MW IMF.  The most typical observed galaxy at the currently sampled redshifts is on the star-forming main sequence.  However, our Galaxy with $sSFR \sim 10^{-11}$ yr$^{-1}$ \citep{licquia2015} is consistent with being quiescent.  Besides, the existing empirical IMFs probe only small selected regions in the MW, which are not guaranteed to represent the average conditions in the Galaxy.  Thus, if it is possible to fix an IMF in a random photometric-template fitting experiment, the local IMF is unlikely to be the preferred option.

\subsection{Monolithic Galaxy Bias} \label{sec:monolithic_bias}

Assuming that the diversity of gas temperatures and compositions can be solved for, the other key approximation is that of galaxies as point sources.  The structural extent and large crowding of stars is neglected in spatially unresolved studies.  So it is possible that the integrated light can be biased towards the brightest stars. In this case, the old stellar populations may be outshined by luminous young O and B-type stars in star-forming galaxies.  Besides, galaxies can have different morphology, from compact cores to complex spiral arms with bars and ellipticals, where, depending on relative locations of different stellar populations, each probably results in a different bias.

Recent studies comparing spatially-resolved and integrated photometric observations discover relative bias in recovered stellar populations.  \cite{gimenezarteaga2023} show that for five compact early galaxies at $5 < z < 9$ this leads to their masses being underestimated by a factor of up to 1.0 dex.  It is found that the young stars born in the central compact region dominate the integrated photometry over the older stars. \cite{sorba2015,sorba2018} showed similar results, with up to 80\% of the spatially-resolved stellar mass missing from their integrated analysis at $0.25 < z < 2.50$.

As most extragalactic observations lack spatial resolution, it is key to quantify the reported bias and factor it into physical properties of galaxy ensembles.  This effect is likely to be the strongest for the most actively star-forming galaxies with blue colors and especially young starbursts.  It is also likely that it is easier to ``conceal'' the stellar mass in the more compact structures.  Therefore, it is necessary to investigate this effect for dusty star-forming galaxies, galaxies with different sSFR and various morphologies at different epochs.  The empirical correction for stellar mass has been derived by \cite{sorba2018} for galaxies at $0.25 < z < 2.50$ based on the sSFR.  The work reported the strongest bias at increasing redshift and high sSFR, but has not investigated the connection to galaxy morphology.

The effect of the combination of the temperature-dependent or else physically-motivated IMF with the integrated photometry bias will perhaps depend on the bottom-lightness of the IMF and the total galaxy luminosity at which the photometry is dominated by the youngest O and B-type stars.  Calculating the total systematic effect is necessary to properly compare the observed physical properties of galaxies with the cosmological predictions.

Finally, the photometric decomposition of main-sequence galaxies into disks and buldges or central cores in \cite{abramson2014} demonstrates that morphological properties of galaxies may be used to further dissect galaxy evolution into its fundamental constituents.  It was shown that the SFMS may be reconstructed with the star-forming galaxy disks alone.  Such interpretation would be consistent with the picture of galaxy evolution here and in \cite{steinhardt2023}, where the core-forming galaxies may be progenitors of the main sequence disks (see \S~\ref{sec:hrdiagram}).

\subsection{Temperature Proxies for IMF} \label{sec:imf_prior}

In this work it was possible to constrain the IMF only for the $\sim 10 \%$ of COSMOS2020 catalog with the widest wavelength coverage and highest S/N, leaving most of the observations unused.  The unused part of the sample demonstrated strong covariances between IMF and other model parameters.  Therefore, additional information in terms of extending the wavelength coverage, probing deeper photometry or setting a prior IMF is required to make use of the available data.

In the framework of finding best-fit model photometry it is common to impose a magnitude-based prior to zoom in on a physically motivated part of the parameter space, which would otherwise suffer from statistical degeneracies \citep{brammer2008}.  Similar approach is commonly performed with the IMF when the MW prior is used.  If there are indications that the ISM changes at higher redshift can drive the changes in mass scale of stellar populations, a prior on the IMF should be set correspondingly to the relevant physical regime.

For setting the IMF, it is necessary to determine the gas temperature in the relevant redshift regime.  It is thought that multiple feedback processes may be responsible for setting the temperature of the star-forming gas in the galaxy \citep{papadopoulos2010}.  However, currently there is no preferred method for predicting this temperature in a galaxy.  Therefore, it is only possible to determine the minimum CMB temperature with certainty.  As it is demonstrated in \cite{steinhardt2022c_templates}, the starting point can be setting the IMF based on the redshift regime, which corresponds to an approximate CMB temperature band.  Although not ideal, this method should almost certainly be more accurate than the assumption of the local IMF.

In absence of theoretical models for determining the gas temperatures at redshift $z > 2$, it may be possible to use the temperature of the cold dust component as a proxy.  The possible correlation between excitation temperature of the gas and cool dust in galaxies on the main sequence \citep{valentino2020} may be used as prior temperature for IMF.  Thus, at redshifts $2 < z < 6$ where the CMB is below the local MW temperature, such proxy for an IMF could be tested for galaxies on the main sequence.  However, as discussed in \S~\ref{sec:hrdiagram}, different feedback processes are likely responsible for galaxies not on the main sequence, such as the starbursts or quiescent galaxies.  Therefore, it is not guaranteed that the same temperature correlations can be expected in those galaxies.

A clear example when it is necessary to use prior temperature knowledge are early galaxies found with the James Webb Space Telescope.  Their redshift regime ($z>10$) corresponds to the most extreme star-forming conditions compared to the nearby galaxies, where even the CMB temperature places the lower temperature limit to over 30 K.  Therefore, the local IMF assumption (equivalent to 20 K) is physically implausible in these galaxies.  However, it is likely not possible to constrain the IMF due to insufficient photometric band coverage in the early observations.  Therefore, it is necessary to ensure that the IMF shape meets the minimum temperature requirement set by the CMB.  Alternatively, the masses can be overestimated by as much as 0.4-1.0 dex \citep{steinhardt2022c_templates}.

\subsection{Impossibly Early Galaxies} \label{sec:early_gals}

In the previous decade, it became possible to test the predictions of the hierarchical assembly of the early galaxies with the advent of such surveys as Cosmic Assembly Near-infrared Deep Extragalactic Legacy Survey (CANDELS; \citealp{grogin2011,koekemoer2011}) and Spitzer Large Area Survey with Hyper-Suprime-Cam (SPLASH) that allowed observations of sizeable samples of galaxies at $4 < z < 8$.  \cite{davidzon2017} and \cite{steinhardt2016}, among others, reported a sharp problem where the observed stellar mass growth is more efficient than allowed in the current cosmological model at high redshift.  It was found that by converting the photometrically-derived stellar masses to halo masses based on local measurements or by matching galaxies luminosities to the available dark matter halos, the inferred halo masses exceeded the theoretical predictions by as much as $\sim 0.8$ dex at $z=10$ \citep{steinhardt2016}.  Similarly, the stellar mass functions appeared to outgrow the halo masses at $z > 3.5$ \citep{davidzon2017}.

More recently, the disagreement between theory and observations has been extended even to higher redshift of $z > 8$ with some early JWST observations (eg., \citealp{labbe2023}).  It has been shown that these massive galaxies have outgrown their dark matter halos several times \citep{boylankolchin2023} making it critical that either the Lambda Cold Dark Matter model \citep{planck2020} 
is incorrect or the interpretation of observations has a fault.  Further studies have either not found the same disagreement \citep{adams2023, mccaffrey2023} or have resolved it by reducing its statistical significance \citep{chen2023} and by using a top-heavier IMF \citep{harikane2023,yung2023}.

It is possible that the increasing disagreement at higher redshift is propagated by the assumptions of the physics of star formation at low redshift.  In this case, possibly the analysis of stellar populations at increasingly higher redshift in \cite{davidzon2017,labbe2023} and other work shows a progressively greater bias due to the bottom-heavy IMF.  As it is argued in \S~\ref{sec:mw_bias}, the conditions for star formation are expected to become increasingly different from the local universe in the regimes of higher CMB temperature and correspondingly higher stellar radiation and cosmic ray feedback at high redshift.  If that is the case, the mass of the stellar populations in the early galaxies has been overestimated.

Although it is not yet possible to constrain an IMF at high redshift due to the lack of wide photometric information or a large ensemble of rest-frame UV and optical galaxy spectra, this work attempts to demonstrate the effect at lower redshift regime where the survey data are more abundant.  It is found that stellar masses even in this already cosmologically ``cold'' regime at $0 < z < 2$ decrease by up to 0.6 dex with the best-fit IMF, with similar results shown in \cite{sneppen2022} in COSMOS2015 galaxies.

With the constraints demonstrated at low redshift, it has been possible to make rough predictions of the IMF at much higher redshift regimes which can place the interpreted stellar masses of galaxies back in agreement with the cosmological predictions.  \cite{steinhardt2022c_templates} showed that by assuming that the temperature of the star-forming gas and respective IMF is set only by the CMB, the stellar mass estimates at $z > 8$ can be placed well within the cosmologically allowed regime.  Although at $z \sim 10$ the CMB temperature ($\sim 30$ K) becomes substantially higher than in the Milky Way ($\sim 20$ K), it is likely that the various feedback processes associated with the star formation and supernovae can set this temperature even higher leading to much lower estimates of the total stellar masses.

Finally, several assumptions have been made commonly to produce estimates of total galaxy masses at all redshifts, including the local IMF, the specific stellar baryon fraction from abundance matching and the stellar to halo mass fraction.  It is argued here that the first assumption is one of the most counter-intuitive based on the expected temperature changes in the early galaxies.  If the change in estimated stellar masses from lower redshift IMF constraints is extrapolated to higher redshift, it can be sufficient to reconcile tension with the theoretical predictions of halo masses.  However, the fate of the stellar mass crisis will only be clearer after also incorporating the bias from spatially non-resolved stellar populations \citep{gimenezarteaga2023}.  Among other assumptions, the stellar-to-halo mass conversion is also tied to the local measurements which do not have to hold at high redshifts and thus also have to be tested.

\section{Conclusion}

In summary, this work presents new physical properties of $\sim 10^5$ well-measured COSMOS2020 galaxies at $0 < z < 2$ for which it is most likely to constrain an IMF \citep{sneppen2022}.  It is found that the galaxies are best fitted with a continuum of IMFs.  Most of these IMFs are bottom-lighter than measured in the Milky Way.  As a result, stellar mass and star formation rate estimates change significantly.  The decrease in $M_\star$ and $SFR$ of stellar populations is indicative of the increasing gas temperature in molecular clouds at higher redshift, by construction of the IMF employed here.  Consequently, this work also derives the IMF parameter as a measure of the IMF shape and the gas temperature, in theory.  Below is a list of the main features and some of the most interesting general implications of this catalog to galaxy evolution identified so far.

\begin{itemize}
    \item \textit{Change in $SFR$ and $M_{\star}$ properties}.  From the perspective of the temperature-dependent IMF, stellar masses and star formation rates of star-forming galaxies reported in the standard COSMOS2020 catalog up to $z < 2$ are overestimated by factors of $\sim1.6-3.5$ and $2.5-70.0$, respectively.
    \item \textit{IMF shape as a probe of gas temperature.}  The changes in physical properties correspond to the overall increase in the temperature of a star-forming gas as a function of redshift inferred form the IMF.  The theoretical interpretation of the IMF shape as the gas temperature probe agrees with the evidence of increasing dust and gas temperature with redshift \citep{valentino2020,magdis2017,magnelli2014}.  The properties of quiescent galaxies exhibit only a small change towards $z \sim 2$, which is consistent with the measurements of dust temperature for quiescent galaxies in \cite{magdis2021}. 
    \item \textit{Sharp downsizing effect}.  The fraction of quiescent galaxies as a function of stellar mass appears to be consistent with the effect of downsizing \citep{juneau2005,cowie1996}.  The change in stellar masses with the best-fit IMF is differential, such that the star-forming stellar mass functions are shifted towards the lower stellar mass, while the masses of the quiescent galaxies mostly remain unchanged.  As a result, there is a larger fraction of high-stellar mass passive galaxies at every studied redshift, which is more sharply consistent with the effect of mass downsizing in comparison with the standard set of properties.
    \item \textit{The $SFR - M_{\star} - T_{IMF}$ relation}.  The best-fit IMF parameter provides new insights about gas temperature on the SFMS and around it.  There is a clear positive gradient of IMF temperatures with sSFR on the main sequence.  In turn, quiescent galaxies have the lowest temperature at all redshifts.  Additionally, the galaxies with the highest temperatures and lowest stellar masses appear to have a relationship with the IMF temperature distinct from that of the typical star-forming galaxies. 
    \item \textit{The $sSFR - T_{IMF}$ relation}. Main-sequence galaxies exhibit a strong $sSFR - T_{IMF}$ coupling.  However, there appear to be two groups of galaxies for which this relation breaks.  The first tail at low temperature corresponds to quiescent galaxies, while the tail at the high temperature is populated with the lowest-mass active galaxies.  Based on the likely compact morphologies \citep{steinhardt2023}, it has been proposed that the last group is in the early phase of galactic evolution that is distinct from typical galaxies on the SFMS.
    
\end{itemize}

This work releases the new catalog and the photometric templates used here for public access.

\begin{acknowledgments}

We would like to thank John Weaver, Gabriel Brammer, Sune Toft, Kate Gould, Clara Gim\'{e}nez-Arteaga and members of the COSMOS Collaboration for help provided with this work and useful discussions.  The Cosmic Dawn Center (DAWN) is funded by the Danish National Research Foundation under grant No. 140.  This research has made use of the SIMBAD database, operated at CDS, Strasbourg, France.  This research has made use of the VizieR catalogue access tool, CDS, Strasbourg, France (DOI: 10.26093/cds/vizier). The original description of the VizieR service was published in 2000, A\&AS 143, 23.  This research has also made use of NASA Astrophysics Data System Bibliographic Services.

This work made use of the photometric data in the COSMOS2020 catalog which, along with the standard physical properties, can be found at \dataset[10.26093/cds/vizier.22580011]{https://doi.org/10.26093/cds/vizier.22580011}.

\end{acknowledgments}


%\vspace{5mm}

\software{\texttt{EAZY} \citep{brammer2008,eazypy2021},
          \texttt{astropy} \citep{astropy2013,astropy2018},
          \texttt{matplotlib} \citep{matplotlib},
          \texttt{numpy} \citep{numpy2020},
          \texttt{scipy} \citep{scipy2020}.
          }
%\facilities{HST(STIS)}


%\software{astropy \citep{2013A&A...558A..33A,2018AJ....156..123A}, 
%          Cloudy \citep{2013RMxAA..49..137F}, 
%          Source Extractor \citep{1996A&AS..117..393B}
%          }



\appendix

\section{Quality cuts \& outliers} \label{sec:app_outliers}

\begin{figure}
    \centering
    \includegraphics[width=0.5\textwidth]{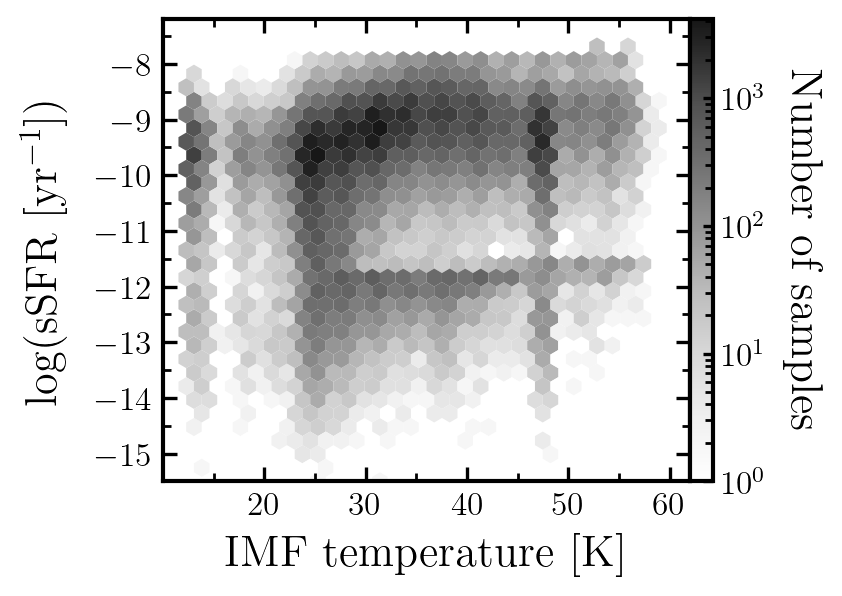}
    \caption{Diagram of sSFR as a function of temperature for galaxies at $0.2 < z < 2.0$.  The grid-like pattern with the vertical stripe at the temperature $47-48$ K and the horizontal stripe at $-12.5 < sSFR < -11.5$ yr$^{-1}$ was flagged as outliers.  They likely originate due to the highest-mass IMF break being covariant with the star formation history.}
    \label{fig:hr_outliers}
\end{figure}

The best-fit SED templates of some samples converged around several very specific parts of the parameter space.  They appear in a grid-like pattern shown in Figure~\ref{fig:hr_outliers} as a vertical stripe at the temperature $47-48$ K and a horizontal stripe at $-12.5 < sSFR < -11.5$ yr$^{-1}$.  In addition, most objects at 18-20 K were part of the outliers as spurious fits with narrow chi-square solutions.  Most of these fits appeared to have the IMF constraint degenerate with the star formation history.  In comparison, the sample of COSMOS2015 objects with variable IMF had a smaller number of apparently similar outliers, although in a different part of the parameter space (cf. \citealp{steinhardt2022_III}).  

These samples were removed from the results demonstrated in this work using the following selection criteria.  They appeared to have distinct mass-to-light ratios $M/L_V$ for their specific star formation rates $SFR/M_\star$, as well as unusually narrow width of the chi-squared solutions $\sigma_T$.  Thus, the following criteria were used to cut the solutions corresponding to: horizontal outliers with $log_{10}(SFR/M_\star) < -11.5$ \& $M/L_V < 0.9~M_{\odot}/L_{\odot}$; vertical outliers with $46 < T_{IMF} < 49$ \& $M/L_V > 0.3$.  These criteria were not optimal, as some viable solutions were cut as a result and some outliers remained in the sample.

\begin{figure*}
    \centering
    \includegraphics[width=0.9\textwidth]{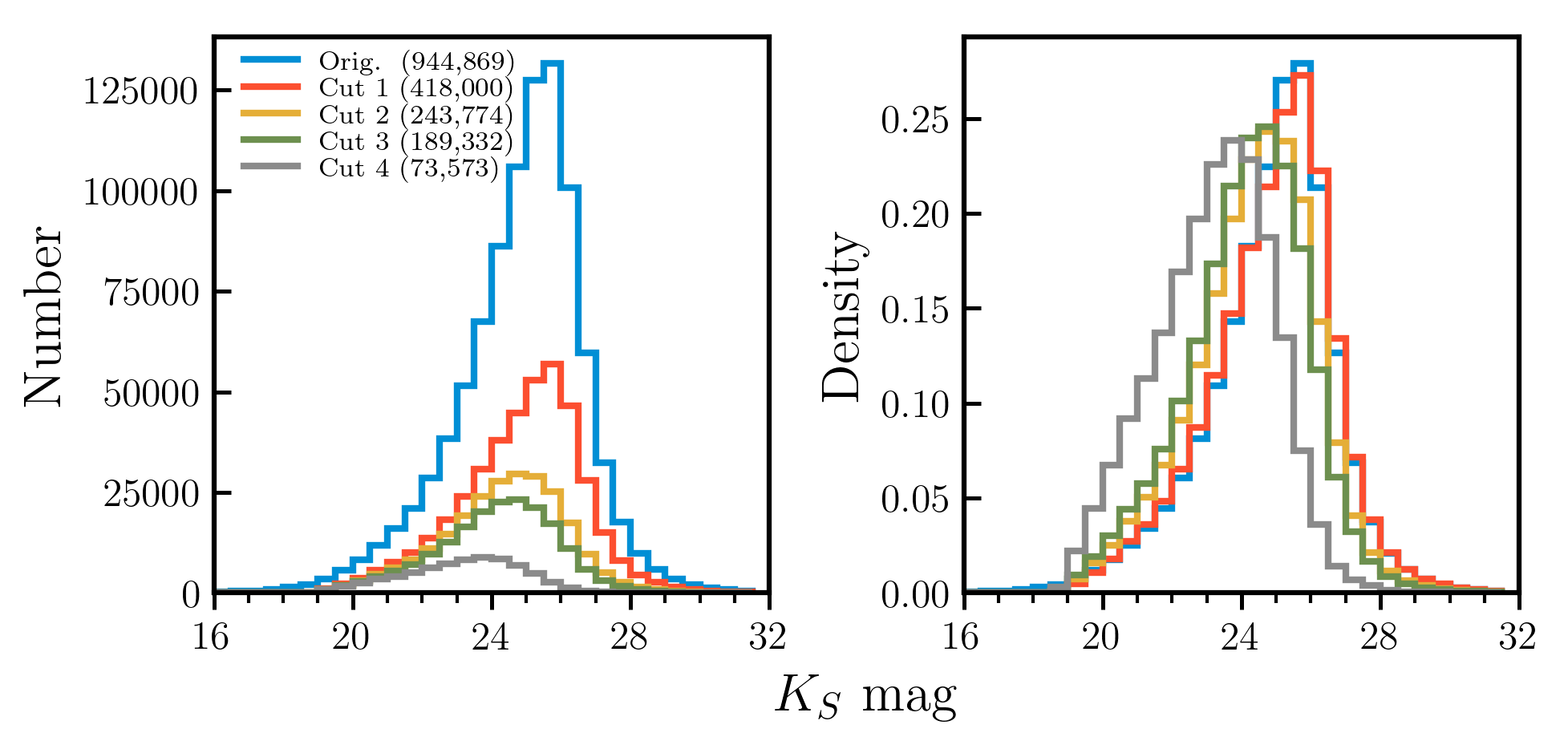}
    \caption{Distributions showing $K_S$ magnitude number histogram (left) and density (right) of the data set after each selection or quality cut.  The sizes of the data set are indicated in the brackets.  The magnitude observable is used to cross-check that the selections do not introduce significant bias in the data.}
    \label{fig:data_cuts}
\end{figure*}

Given that this catalog cuts the original data after several selections, and identifies outliers in the best-fit results, it is important to cross-check the data consistency at every stage.  Figure~\ref{fig:data_cuts} shows the distributions of $K_S$ magnitude after each of the following selections (the sizes of the data sets are given in brackets):

\begin{itemize}
    \item \textit{Cut 1 (418,000)}: ``ultradeep'' stripes of the UltraVISTA photometry. Removing the sources that were biased by the bright neighbors, appeared at the image boundaries, were saturated or failed at execution, as flagged by \texttt{SExtractor}.  The selection reduced the survey area from $\sim3.4$ to $0.9$ deg$^2$.

    \item \textit{Cut 2 (243,774)}: valid IMF solutions, as defined by chi-squared values in the sampled temperature range $10 < T < 60$ K.

    \item \textit{Cut 3 (189,332)}: removing solutions flagged as ``outlier'' (appeared as a grid in Figure~\ref{fig:hr_outliers}), based on the mass-to-light ratio cuts described above.

    \item \textit{(Optional) Cut 4 (73,573)}: this sample includes only the solutions where the uncertainty could be estimated based on the chi-squared metric using the minimum variance bound estimator.  This sample was used to demonstrate most results in this work and is optional. The histograms in Figure~\ref{fig:data_cuts} show that this cut removes a fraction of faint sources.

\end{itemize}

The histograms of the $K_S$ magnitude do not differ significantly and thus indicate that there are no major biases introduced by the selections made in the data. However, the last cut is not advised, as it appears to remove a fraction of $K_S$-faint source.

\section{Changes in Stellar Mass Functions} \label{sec:appendix_smf}

\begin{figure*}[ht]
    \centering
    \includegraphics[width=\textwidth]{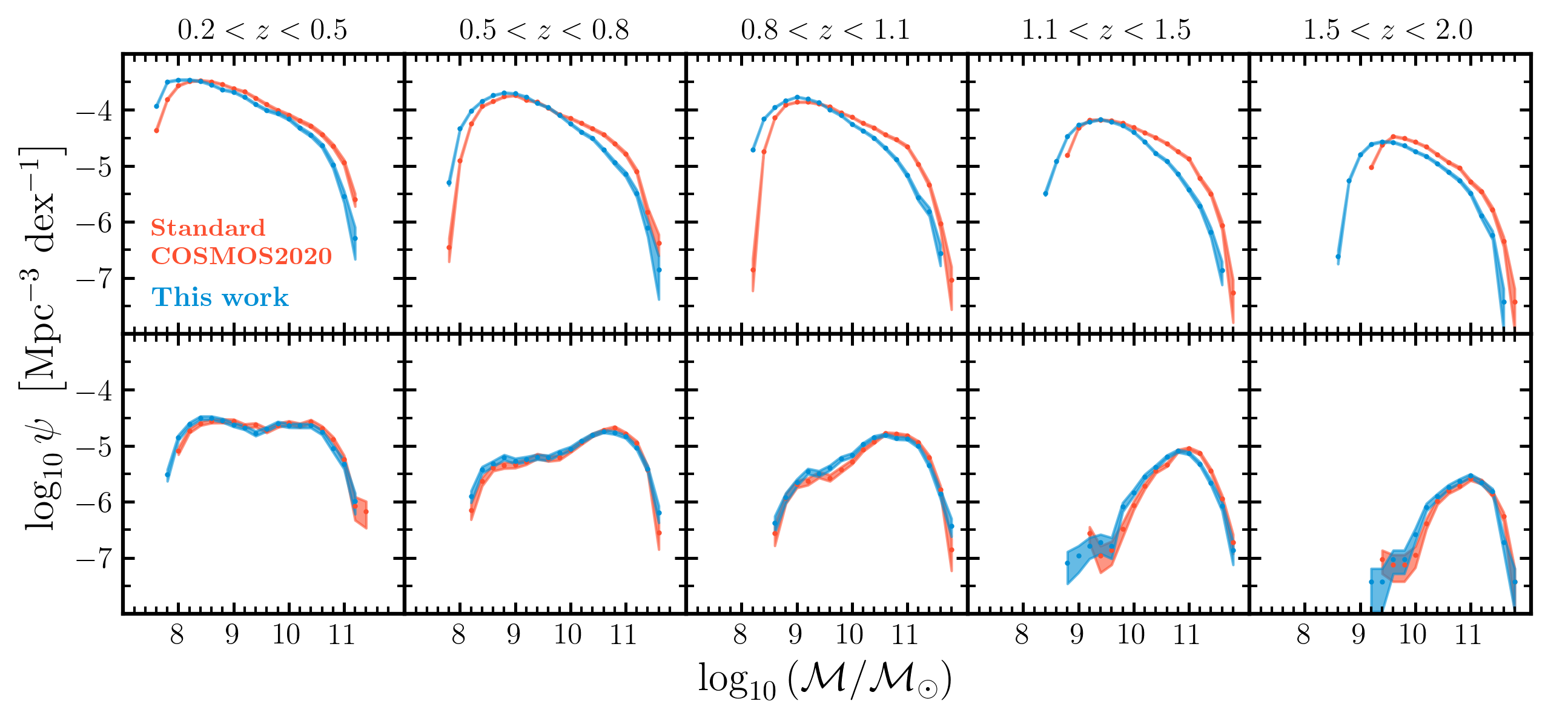}
    \caption{Stellar mass functions for star-forming (top) and quiescent (bottom) galaxies at different redshift windows in $0.2 < z < 2.0$.  The standard stellar mass has been converted from Chabrier to Kroupa IMF to match the properties in the value-addded catalog.  Most masses of active galaxies have been shifted towards lower values as a result of using the best-fit IMF.  On the other hand, the masses of the passive galaxies remain similar to the properties with the Milky Way IMF.  This differential change in stellar mass results in a systematic change of the quiescent galaxy fraction as a function of the mass, which implies a sharper effect of mass downsizing (see Figure~\ref{fig:smf}).}
    \label{fig:smfs_imf_w22}
\end{figure*}

This appendix shows the differences between the stellar mas functions of active and passive galaxies computed using the standard stellar masses from COSMOS2020 and the masses from the value-added catalog with the best-fit IMFs.  These functions are plotted in Figure~\ref{fig:smfs_imf_w22} for star-forming (top row) and quiescent (bottom row) galaxies separately.

\section{Catalog Table} \label{sec:catalog_table}

\begingroup
\renewcommand{\arraystretch}{0.8}

\begin{table}
\begin{threeparttable}[b]
\centering
    \caption{Description of the column labels in the value-added catalog. Rows 1-3: ID and the sky coordinates are taken from the COSMOS2020 \texttt{CLASSIC} catalog. Rows 4-55: standard \texttt{EAZY}-generated best-fit output. Rows 56-59: best-fit IMF temperature (i.e., bottom-lightness of the Kroupa IMF) with its uncertainties and outlier flags.}
    \label{tab:catalog_table}
    
    \begin{tabular}{cccl}
        \hline\hline
        Num & Label & Unit & Description  \\ \hline

        1  & ID                       & ---  &  COSMOS2020 \texttt{CLASSIC} catalogue ID  \\
        2  & ALPHA\_J2000              & deg  &  Right ascension (J2000)   \\
        3  & DELTA\_J2000              & deg  &  Declination (J2000)   \\
        4  & z\_phot                   & ---  &  \texttt{EAZY} maximum likelihood photo-z  \\
        5  & z\_phot\_chi2              & ---  &  chi-squared at z\_phot  \\
        6  & z\_phot\_risk              & ---  &  risk parameter \citep{tanaka2018} at z\_phot  \\
        7  & z\_min\_risk               & ---  &  z\_phot where risk  parameter z\_phot\_risk is minimised  \\
        8  & min\_risk                 & ---  &  risk at z\_min\_risk   \\
        9  & z\_raw\_chi2               & ---  &  photo-z where z\_phot\_chi2 is minimised, without priors   \\
        10 & raw\_chi2                 & ---  &  chi-squared at z\_raw\_chi2   \\
        11 & z$_{\rm xxx}$\tnote{$\star$}            & ---  &  percentiles of photo-z   \\
        12 & nusefilt                 & ---  &  number of filters used for photo-z   \\
        13 & lc\_min                   & 0.1nm  &  minimum effective  wavelength of valid  filters  \\
        14 & lc\_max                   & 0.1nm  &  maximum effective  wavelength of valid  filters  \\
        15 & restU                    & $\mu$Jy  &  rest-frame U-band flux density   \\
        16 & restU\_err                & $\mu$Jy  &  rest-frame U-band flux density uncertainty   \\
        17 & restB                    & $\mu$Jy  &  rest-frame B-band flux density   \\
        18 & restB\_err                & $\mu$Jy  &  rest-frame B-band flux density uncertainty   \\
        19 & restV                    & $\mu$Jy  &  rest-frame V-band flux density   \\
        20 & restV\_err                & $\mu$Jy  &  rest-frame V-band flux density uncertainty   \\
        21 & restJ                    & $\mu$Jy  &  rest-frame J-band flux density   \\
        22 & restJ\_err                & $\mu$Jy  &  rest-frame J-band flux density uncertainty   \\
        23 & dL                       & [Mpc]  &  $\log_{10}$ luminosity distance at z\_phot  \\
        24 & mass                     & [$M_{\odot}$]  &  $\log_{10} M_{star}$   \\
        25 & sfr                      & [$M_{\odot}$ yr$^{-1}$]  &  $\log_{10} {\rm SFR}$   \\
        26 & ssfr                     & [yr$^{-1}$]  &  $\log_{10} {\rm sSFR}$  \\
        27 & Lv                       & [$L_{\odot}$]  &  $\log_{10} L$ (V-band)  \\
        28 & LIR                      & [$L_{\odot}$]  &  $\log_{10} L$ (8-1000 $\mu$m)  \\
        29 & energy\_abs               & [$L_{\odot}$]  &  $\log_{10} E_{\rm absorbed}$ associated with $A_V$   \\
        30 & Lu                       & [$L_{\odot}$]  &  $\log_{10} L$ (U band)   \\
        31 & Lj                       & [$L_{\odot}$]  &  $\log_{10} L$ (J band)   \\
        32 & L1400                    & [$L_{\odot}$]  &  $\log_{10} L$ in 200 $\text{\AA}$-wide top-hat filter (at 1400 $\text{\AA}$)   \\
        33 & L2800                    & [$L_{\odot}$]  &  $\log_{10} L$ in 200 $\text{\AA}$-wide top-hat filter (at 2800 $\text{\AA}$)   \\
        34 & LHa                      & [$L_{\odot}$]  &  $\log_{10} L$(H$\alpha$)  \\
        35 & LOIII                    & [$L_{\odot}$]  &  $\log_{10} L$(\ion{O}{3})  \\
        36 & LHb                      & [$L_{\odot}$]  &  $\log_{10} L$(H$\beta$)  \\
        37 & LOII                     & [$L_{\odot}$]  &  $\log_{10} L$(\ion{O}{2})  \\
        38 & MLv                      & $M_{\odot}$ $L_{\odot}^{-1}$  &  mass-to-light ratio (V band)  \\
        39 & Av                       & mag  &  extinction (V band)   \\
        40 & lwAgeV                   & Gyr  &  light-weighted age (V band)   \\
        41 & mass\_p$_{\rm xxx}$\tnote{$\star$}       & [$M_{\odot}$]  &  percentiles of $\log_{10} M_{star}$  \\
        42 & sfr\_p$_{\rm xxx}$\tnote{$\star$}        & [$M_{\odot}$ yr$^{-1}$]  &  percentiles of $\log_{10} {\rm SFR}$   \\
        43 & Lv\_p$_{\rm xxx}$\tnote{$\star$}         & [$L_{\odot}$]  &  percentiles of $\log_{10} L$ (V band) \\
        44 & LIR\_p$_{\rm xxx}$\tnote{$\star$}        & [$L_{\odot}$]  &  percentiles of $\log_{10} L$ (8-1000 $\mu$m)   \\
        45 & energy\_abs\_p$_{\rm xxx}$\tnote{$\star$} & [$L_{\odot}$]  &  percentiles of $\log_{10} E_{\rm absorbed}$ associated with $A_V$   \\
        46 & Lu\_p$_{\rm xxx}$\tnote{$\star$}         & [$L_{\odot}$]  &  percentiles of $\log_{10} L$ (U band)   \\
        47 & Lj\_p$_{\rm xxx}$\tnote{$\star$}         & [$L_{\odot}$]  &  percentiles of $\log_{10} L$ (J band)   \\
        48 & L1400\_p$_{\rm xxx}$\tnote{$\star$}      & [$L_{\odot}$]  &  percentiles of $\log_{10} L$ in 200 $\text{\AA}$-wide top-hat filter (at 1400 $\text{\AA}$)  \\
        49 & L2800\_p$_{\rm xxx}$\tnote{$\star$}      & [$L_{\odot}$]  &  percentiles of $\log_{10} L$ in 200 $\text{\AA}$-wide top-hat filter (at 2800 $\text{\AA}$)  \\
        50 & LHa\_p$_{\rm xxx}$\tnote{$\star$}        & [$L_{\odot}$]  &  percentiles of $\log_{10} L$(H$\alpha$) \\
        51 & LOIII\_p$_{\rm xxx}$\tnote{$\star$}      & [$L_{\odot}$]  &  percentiles of $\log_{10} L$(\ion{O}{3}) \\
        52 & LHb\_p$_{\rm xxx}$\tnote{$\star$}        & [$L_{\odot}$]  &  percentiles of $\log_{10} L$(H$\beta$)  \\
        53 & LOII\_p$_{\rm xxx}$\tnote{$\star$}       & [$L_{\odot}$]  &  percentiles of $\log_{10} L$(\ion{O}{2})  \\
        54 & ssfr\_p$_{\rm xxx}$\tnote{$\star$}       & [yr$^{-1}$]  &  percentiles of $\log_{10} {\rm sSFR}$  \\
        55 & Av\_p$_{\rm xxx}$\tnote{$\star$}         & mag  &  percentiles of extinction (V band)  \\
        56 & t\_imf                    & K  &  $T_{\rm IMF}$, maximum-likelihood temperature of Kroupa IMF \citep{jermyn2018}  \\
        57 & t\_imf\_siglo              & K  &  $T_{\sigma_-}$, uncertainty in  best-fit temperature of Kroupa IMF   \\
        58 & t\_imf\_sigup              & K  &  $T_{\sigma_+}$, uncertainty in  best-fit temperature of Kroupa IMF   \\
        59 & outlier                  & ---  &  bad local solutions  that appear as a grid of $T_{\rm IMF}$ and sSFR (see \S~\ref{sec:app_outliers})   \\ \hline
    \end{tabular}
    \begin{tablenotes}
       \item [1] ${\rm xxx}$ is one of [025, 160, 500, 840, 975] percentiles, i.e. [2.5\%, 16\%, 50\%, 84\%, 97.5\%], of a PDF of a quantity.
    \end{tablenotes}
    \tablecomments{Table~\ref{tab:catalog_table} is published in its entirety in the machine-readable format and made available online.  A portion is shown here for guidance regarding its form and content.}
\end{threeparttable}
\end{table}

\endgroup


\bibliography{references}{}
\bibliographystyle{aasjournal}



\end{document}